\documentclass[aps,prl,reprint,groupedaddress]{revtex4-2}

\usepackage{amsmath,graphicx,color}
\usepackage{bm}
\begin{document}

\title{Distinguishing finite momentum superconducting pairing states with two-electron photoemission spectroscopy}
\author{Fahad Mahmood$^{1,2}$, Thomas Devereaux$^{3,4}$, Peter Abbamonte$^{1,2}$, and Dirk K. Morr$^{5}$}
\affiliation{$^{1}$ Department of Physics, University of Illinois at Urbana-Champaign,Urbana,IL 61801, USA}
\affiliation{$^{2}$ Materials Research Laboratory, University of Illinois at Urbana-Champaign,Urbana,IL 61801, USA}
\affiliation{$^{3}$ Department of Materials Science and Engineering, Stanford University, Stanford, CA, USA}
\affiliation{$^{4}$ Stanford Institute for Materials and Energy Sciences, SLAC National Accelerator Laboratory, Menlo Park, CA, USA}
\affiliation{$^{5}$ University of Illinois at Chicago, Chicago, IL 60607, USA}
\date{\today}

\begin{abstract}
We show theoretically that double photoemission (2$e$-ARPES) may be used to identify the pairing state in superconductors in which the Cooper pairs have a nonzero center-of-mass momentum, ${\bf q}_{cm}$. We theoretically evaluate the 2$e$ ARPES counting rate, $P^{(2)}$, for the cases of a $d_{x^2-y^2}$-wave superconductor, a pair-density-wave (PDW) phase, and a Fulde-Ferrel-Larkin-Ovchinnikov (FFLO) phase. We show that $P^{(2)}$  provides direct insight into the center-of-mass momentum and spin state of the superconducting condensate, and thus can distinguish between these three different superconducting pairing states. In addition, $P^{(2)}$ can be used to map out the momentum dependence of the superconducting order parameter. Our results identify 2$e$-ARPES as an ideal tool for identifying and probing ${\bf q}_{cm} \neq 0$ superconducting pairing states in superconductors.
\end{abstract}
\maketitle

{\it Introduction} Identifying the pairing symmetries of unconventional superconductors has remained one of the most important and fundamental challenges in quantum materials research. Its difficulty arises from the absence of two-particle spectroscopies that directly probe the properties of the Cooper pair wave-function, which determine the spin structure and momentum dependence of the superconducting order parameter. Single-particle spectroscopies such as tunneling \cite{spectral2012} or angle-resolved photoemission spectroscopy (ARPES) \cite{damascelli2003} can only measure the magnitude of the superconducting order parameter (or the gap) but not its phase, while macroscopic Josephson interference measurements can probe its phase, but only if the order parameter is spatially uniform and suitable junctions can be prepared \cite{DVH1995,Tsuei2000}.
The difficulties are even more acute for superconductors in which the Cooper pairs possess a nonzero center-of-mass momentum, ${\bf q}_{cm}$, such as the Fulde-Ferrel-Larkin-Ovchinnikov (FFLO) phase \cite{Fulde1964,Larkin1965} or the predicted pair density wave (PDW) \cite{berg2009}, in which the superconducting order parameter is modulated in real space.

In this article we demonstrate that two electron coincidence spectroscopy (2$e$-ARPES), in which the absorption of a single photon leads to the emission of two coincident photo-electrons \cite{Berakdar_theory_1998}, can directly reveal the microscopic character of finite-momentum pairing states in superconductors. The experimental 2$e$-ARPES signal, the photo-electron counting rate $P^{(2)}$, is the probability per unit time that a single photon leads to the emission of a correlated pair of photo-electrons with defined energy, momentum, and spin, as measured by two separate detectors. We show theoretically that the $d_{x^2-y^2}$-wave superconducting, FFLO and PDW phases have distinct spectroscopic signatures in $P^{(2)}$, which is related to a two-particle spectral function \cite{napitu_two-particle_2010} that is directly sensitive to the  center-of-mass momentum and spin state of the Cooper pair wave-function. 2$e$-ARPES is therefore a promising technique for identifying and studying spatially modulated superconductors generally.

\begin{figure}[ht]
\centering
\includegraphics[width=8.8cm]{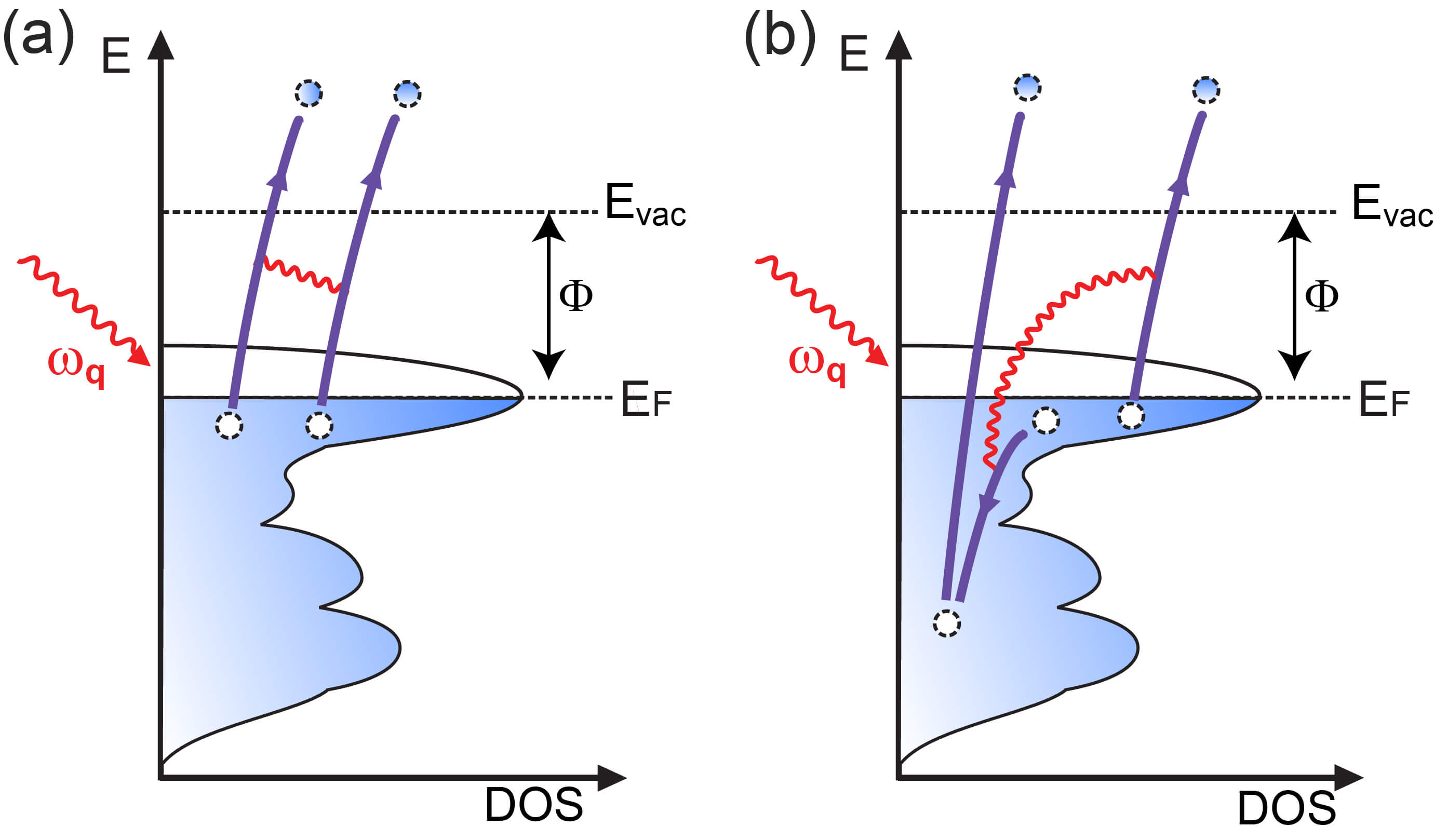}
\caption{Schematic representation of the two distinct 2$e$ ARPES processes involving the absorption of a single photon, and the ejection  of two photo-electrons: (a) The incident photon excites a valence electron into a free photo-electron state with $E>E_{vac}$ which in turn ejects a second valence electron through an EELS-like scattering event. Here, $E_{vac} = E_F + \Phi$ where $E_F$ is the Fermi energy and $\Phi$ is the work function, (b) The incident photon excites an electron from a core level which is then filled by a valence electron, emitting a second valence electron though an Auger process.}
\label{fig:Fig1ex}
\end{figure}

There are two distinct processes that can cause a single photon to lead to the ejection of a correlated pair of electrons [Fig.~\ref{fig:Fig1ex}(a),(b)]  \cite{Berakdar_theory_1998,chiang_laser-based_2020,Fominykh_PRL_2002,Schumann_2007,schumann_surface_2011,schumann_electron_2012,Herrmann_1998}. In the first, the photon is absorbed and excites a valence band electron into a free photo-electron state, which subsequently ejects a second valence electron via an electron energy-loss (EELS)-like scattering event [Fig.~\ref{fig:Fig1ex}(a)]. In the second process, the first photo-electron is excited from a core-level, which is subsequently filled by a valence electron, leading to the emission of a second valence electron through an Auger process [Fig.~\ref{fig:Fig1ex}(b)]. While both processes lead to a very similar energy, momentum and spin dependence of $P^{(2)}$ [see Supplemental Material (SM) Secs. 1 and 2], the use of lower photon energy, laser based XUV sources will not allow 2$e$-ARPES experiments to directly probe core states, rather rendering them more sensitive to valence band effects. We thus restrict our theoretical analysis to first type of process, shown in Fig.~\ref{fig:Fig1ex}(a).

{\it Theoretical Formalism} We compute the 2$e$-ARPES photo-electron counting rate $P^{(2)}$ in the sudden approximation whereby we neglect relaxation pathways during the photo-electron emission process and work with plane wave electrons at the detector and valence electron states in a sample \cite{Pavlyukh2015}.
As mentioned above, we focus on the two-step process shown in Fig.~\ref{fig:Fig1ex}(a) involving the emission of a first photo-electron upon absorption of a photon, and the subsequent scattering (EELS-like) process between the emitted photo-electron and a conduction electron, which leads to the emission of a second photo-electron. We assume that scattering process between the photo-electron and conduction electron is mediated by a (screened) Coulomb interaction.  This entire process is then described by the Hamiltonian
\begin{align}
    H_{sc} & = \sum_{{\bf k,q},\sigma,\nu} \gamma_{\nu}({\bf q}) d^\dagger_{{\bf k+q},\sigma} c_{{\bf k},\sigma} \left( a_{{\bf q},\nu}  + a^\dagger_{-{\bf q},\nu}\right)  \nonumber \\
    & + \sum_{{\bf k,p,q},\alpha, \beta} V({\bf q}) d^\dagger_{{\bf k}+{\bf q},\alpha} d^\dagger_{{\bf p}-{\bf q},\beta} d_{{\bf p},\beta} c_{{\bf k},\alpha}  + h.c.
\end{align}
Here, $\gamma_{\nu}({\bf q})$ is the effective electron-photon dipole interaction, $d^\dagger_{{\bf k},\sigma} (c_{{\bf k},\sigma})$ creates (destroys) a photo-electron (conduction electron) with momentum ${\bf k}$ and spin $\sigma$, and $V({\bf q}) \sim [{\bf q}^2 + \kappa^{2}]^{-1}$ is the Fourier transform of the (screened) Coulomb interaction, with $\kappa^{-1}$ being the screening length. Moreover, since the photon momentum is much smaller than typical fermionic momenta, we set it equal to zero.
As the out-of-plane momentum is not conserved upon absorption of the photon,  we take $\gamma_{\nu}({\bf q}) = \gamma_0$ to be independent of the in-plane momentum.

\begin{figure}[ht]
\centering
\includegraphics[width=8.5cm]{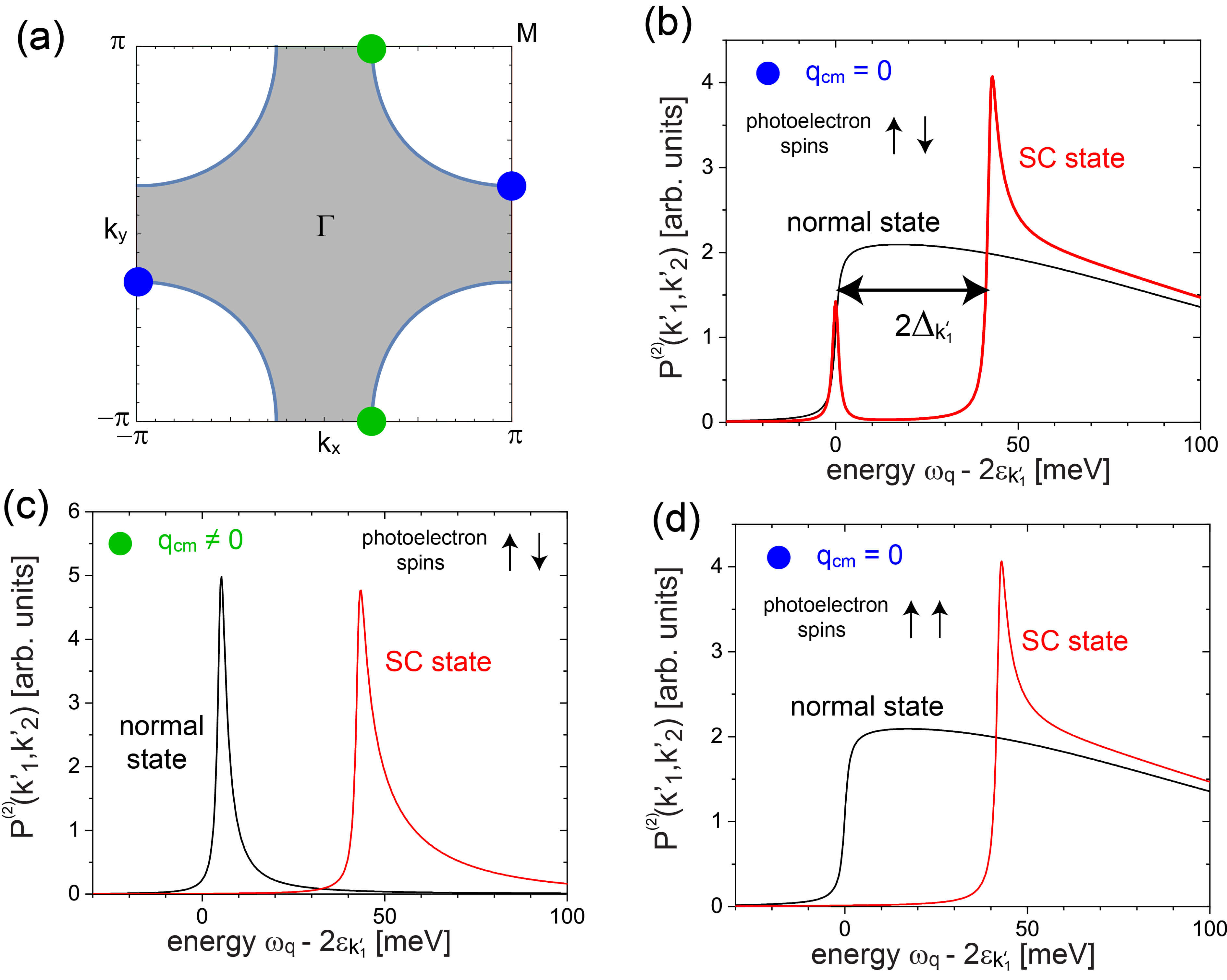}
\caption{(a) Fermi surface of the cuprate superconductors. (b) $P^{(2)}$ in a $d_{x^2-y^2}$-wave superconductors for two photo-electrons with opposite momenta ${\bf k}_2^\prime = -{\bf k}_1^\prime$, as indicated by the set of filled blue circles in (a), and opposite spins, $\sigma_1^\prime \not = \sigma_2^\prime$. $P^{(2)}$ for photo-electrons (c) with momenta indicated by green circles and opposite spins, and (d) with momenta indicated by blue circles and equal spins.}
\label{fig:Fig2}
\end{figure}

The initial and final states of the entire system, $|\Psi_a \rangle$ and $| \Psi_b \rangle$ respectively, are described by \begin{align}
    | \Psi_a \rangle & = | \Phi_a \rangle | 1_{{\bf q}, \lambda} \rangle_p | 0 \rangle_{pe} \nonumber \\
    | \Psi_b \rangle & = | \Phi_b \rangle | 0 \rangle_p | 1_{{\bf k}_1^\prime, \sigma^\prime_1} 1_{{\bf k}_2^\prime, \sigma^\prime_2} \rangle_{pe} \ .
\end{align}
Here, $|1_{{\bf q} \lambda} \rangle_p$ describes the initial photon state containing one photon with momentum ${\bf q}$ and polarization $\lambda$, and $| 1_{{\bf k}_1^\prime, \sigma^\prime_1} 1_{{\bf k}_2^\prime, \sigma^\prime_2} \rangle_{pe}$ represents the final photo-electron state containing two photo-electrons with momenta ${\bf k}_{1,2}^\prime$ and spin $\sigma^\prime_{1,2}$. The initial and final states of the superconductor are described by $| \Phi_{a,b} \rangle$, respectively. The 2$e$-ARPES signal, which depends on the two photo-electron momenta and spin projections, is then computed via
\begin{equation}
 P^{(2)}({\bf k}_1^\prime, \sigma^\prime_1, {\bf k}_2^\prime, \sigma^\prime_2 ) = \frac{1}{Z} \sum_{a,b} \frac{e^{-\beta E_a}}{\Delta t} \left| \langle  \Psi_b | {\hat S}(\infty,-\infty) | \Psi_a \rangle \right|^2
\end{equation}
where $Z$ is the partition function, the sum runs over all states $|\Phi_{a,b} \rangle$ of the superconductor, $\Delta t$ is the time over which the photon beam is incident in the superconductor, and ${\hat S}$ is the $S$-matrix that we expand to second order in $H_{sc}$. The detailed derivation of $P^{(2)}$ for a uniform  $d_{x^2-y^2}$-wave superconductor, the PDW and the FFLO phases is carried out in SM Sec.~1. While we consider for concreteness a cuprate-like Fermi surface, as shown in Fig.~\ref{fig:Fig2}(a), our results shown below are quite general and applicable to a wide variety of superconductors with varying Fermi surface structure.

{\it Results}
We begin by discussing the case of a uniform, spin-singlet $d_{x^2-y^2}$-wave superconductor (band parameters are given in SM Sec.~1) in which the Cooper pairs possess a zero center of mass momentum. For $P^{(2)}$ to directly probe the superconducting condensate, we need to require that the two photo-electrons also have a zero center-of-mass momentum, i.e., ${\bf k}_2^\prime = -{\bf k}_1^\prime$, and opposite spins, i.e., $\sigma^\prime_2 \not = \sigma^\prime_1$. In this case, we obtain $P^{(2)} =  P_{SC}^{(2)} +  P_{2cp}^{(2)}$ where (at $T=0$)
\begin{align}
    P_{SC}^{(2)}
    & = 2 \pi \delta(\omega_q - 2 \varepsilon_{{\bf k}_1^\prime})  \left| \sum_{\bf k}
    \frac{ \gamma_0 V\left({\bf k} - {\bf k}_1^\prime \right)}{\omega_q - E_{\bf k} - \varepsilon_{\bf k}+i \delta} \frac{\Delta_{\bf k}}{2 E_{\bf k} }\right|^2 \nonumber \\
    P_{2cp}^{(2)} & = 2 \pi \sum_{\bf k} \left| \frac{ \gamma_0 V\left({\bf k} - {\bf k}_1^\prime\right) v_{\bf k}^2}{\omega_q + E_{\bf k} - \varepsilon_{\bf k}+i \delta}  \right|^2 \delta(\omega_q - 2 \varepsilon_{{\bf k}_1^\prime} -2  E_{\bf k}) \ ,
    \label{eq:p2}
\end{align}
where $E_{\bf k} = \sqrt{\xi_{\bf k}^2 + \Delta_{\bf k}^2}$ ($\xi_{\bf k}$) is the conduction electron dispersion in the superconducting (normal) state, $v^2_{\bf k}=\left[1 -\xi_{\bf k}/E_{\bf k}\right]/2$, $\omega_q$ is the incident photon energy, and $\varepsilon_{{\bf k}_1^\prime}$ is the sum of the kinetic energy and work function of a photo-electron. $\Delta \omega = \omega_q - 2 \varepsilon_{{\bf k}_1^\prime}$ represents the excess energy of the photon over the energies of the two photo-electrons. The first term, $P_{SC}^{(2)}$,  directly reflects the existence of a superconducting condensate, as described by $\Delta_{\bf k}$, and arises from the breaking and subsequent creation of a Cooper pair. This term vanishes in the normal state, and is absent when the two detected photo-electrons do not possess the same center of mass momentum, or spin structure as the superconducting condensate. As such, the photo-electron pairs that contribute to $P_{SC}^{(2)}$ reside in an entangled state and are therefore Einstein-Podolsky-Rosen (EPR) pairs. Note that the momentum dependence of the Coulomb interaction plays a crucial role in observing a non-zero $P_{SC}^{(2)}$ in a $d_{x^2-y^2}$-wave superconductor since for a momentum independent $V({\bf q})$, $P_{SC}^{(2)}$ vanishes identically due to the symmetry of the $d_{x^2-y^2}$-wave order parameter. In contrast, the photo-electron pairs that contribute to the second term, $P_{2cp}^{(2)}$, arise from the breaking of two Cooper pairs. As $P_{2cp}^{(2)}$ is weighted by the particle-like coherence factors of the broken Cooper pairs, i.e., $(v^2_{\bf k})^2$, it does not vanish in the normal state.

In Fig.~\ref{fig:Fig2}(b) we present $P^{(2)}$ in the normal and superconducting state for opposite photo-electron momenta ${\bf k}_2^\prime \not = -{\bf k}_1^\prime$ near the anti-nodal points [indicated by the filled blue circles in Fig.~\ref{fig:Fig2}(a)]. In the normal state, $P^{(2)}$ shows an onset at $\Delta \omega=0$, as conduction electrons can be excited from the filled Fermi sea for $\Delta \omega \geq 0$.
In contrast, in the superconducting state, $P^{(2)}$ exhibits two distinct features. The first one is a peak at $\Delta \omega=0$, previously identified in Ref. \cite{Berakdar_SC_2003}, arising from $P_{SC}^{(2)}$ in Eq.(\ref{eq:p2}) that is a direct signature of the superconducting condensate, as discussed above. The second feature is a continuum, described by $P_{2cp}^{(2)}$, with onset energy $\Delta \omega_c \approx 2\Delta_{{\bf k}_1^\prime}$ (we will refer to this contribution as the 2CP continuum). The latter immediately reveals that $P_{2cp}^{(2)}$ reflects the measurement of 2 photo-electrons arising from the breaking of 2 Cooper pairs, requiring an energy of $2\Delta_{{\bf k}^\prime_1}$. That the gap between the condensate peak at $\Delta \omega=0$ and the continuum is indeed $2\Delta_{{\bf k}_1^\prime}$ is a direct consequence of the momentum dependence of the Coulomb interaction, $V({\bf q})$, which suppresses large momentum transfers during the scattering process. As a result, the main contribution to $P_{2cp}^{(2)}$ arises from those momentum states ${\bf k}$ along the Fermi surface with ${\bf k} \approx \pm {\bf k}_1^\prime$. As $\Delta \omega_c \approx 2\Delta_{{\bf k}_1^\prime}$, it decreases as one moves from the anti-nodal to the nodal points along the Fermi surface (see SM Sec.~3), allowing one to map out the momentum dependence of the superconducting gap along the Fermi surface, in a similar manner to conventional ARPES experiments \cite{damascelli2003}. Finally, we note that our results are qualitatively robust against changes in the screening length $\kappa^{-1}$ of the Coulomb interaction (SM Sec.~4).

A qualitatively new feature of 2$e$-ARPES is that it can be used to identify the center-of-mass momentum of the Cooper pairs, ${\bf q}_{cm}$. To demonstrate this, we plot in  Fig.~\ref{fig:Fig2}(c) $P^{(2)}$ for photo-electron momenta indicated by filled green circles in Fig.~\ref{fig:Fig2}(a). While each of these momenta by itself is symmetry-related to the momentum indicated by blue circles in Fig.~\ref{fig:Fig2}(a), their sum (i.e., their center-of-mass momentum) is non-vanishing, ${\bf q}_{cm} \not = 0$. As such, $P^{(2)}$ for these two momenta does not exhibit a zero-energy peak [see Fig.~\ref{fig:Fig2}(c)] as the condensate possesses ${\bf q}_{cm}=0$. In contrast, the onset energy for the continuum, $\Delta \omega_c$ is still located at the same energy $2\Delta_{{\bf k}_1^\prime}$ as in Fig.~\ref{fig:Fig2}(b), as it arises from the breaking of two Cooper pairs.

Further, $P^{(2)}$ even reveals the spin-state of the Cooper pairs. In Fig.~\ref{fig:Fig2}(d) we present $P^{(2)}$ for two photo-electrons with the same momenta as in Fig.~\ref{fig:Fig2}(b) (filled blue circles), but possessing equal spins. In this case, $P^{(2)}$ does not exhibit a zero-energy peak (i.e., $P^{(2)}_{sc} \equiv 0$), as the electrons in a Cooper pair form a spin-singlet state. Thus only a measurement of photo-electrons that are in opposite spin states will exhibit a zero-energy peak in $P^{(2)}$. In contrast, the continuum in $P_{2cp}^{(2)}$ is the same for equal and opposite spin states of the photo-electrons, as it arises from the breaking of two Cooper pairs. These results demonstrate that 2$e$-ARPES experiments provide unprecedented insight into the center-of-mass momentum and spin state of the superconducting condensate, as well as the momentum dependence of the superconducting order parameter.

\begin{figure}[ht]
\centering
\includegraphics[width=8.5cm]{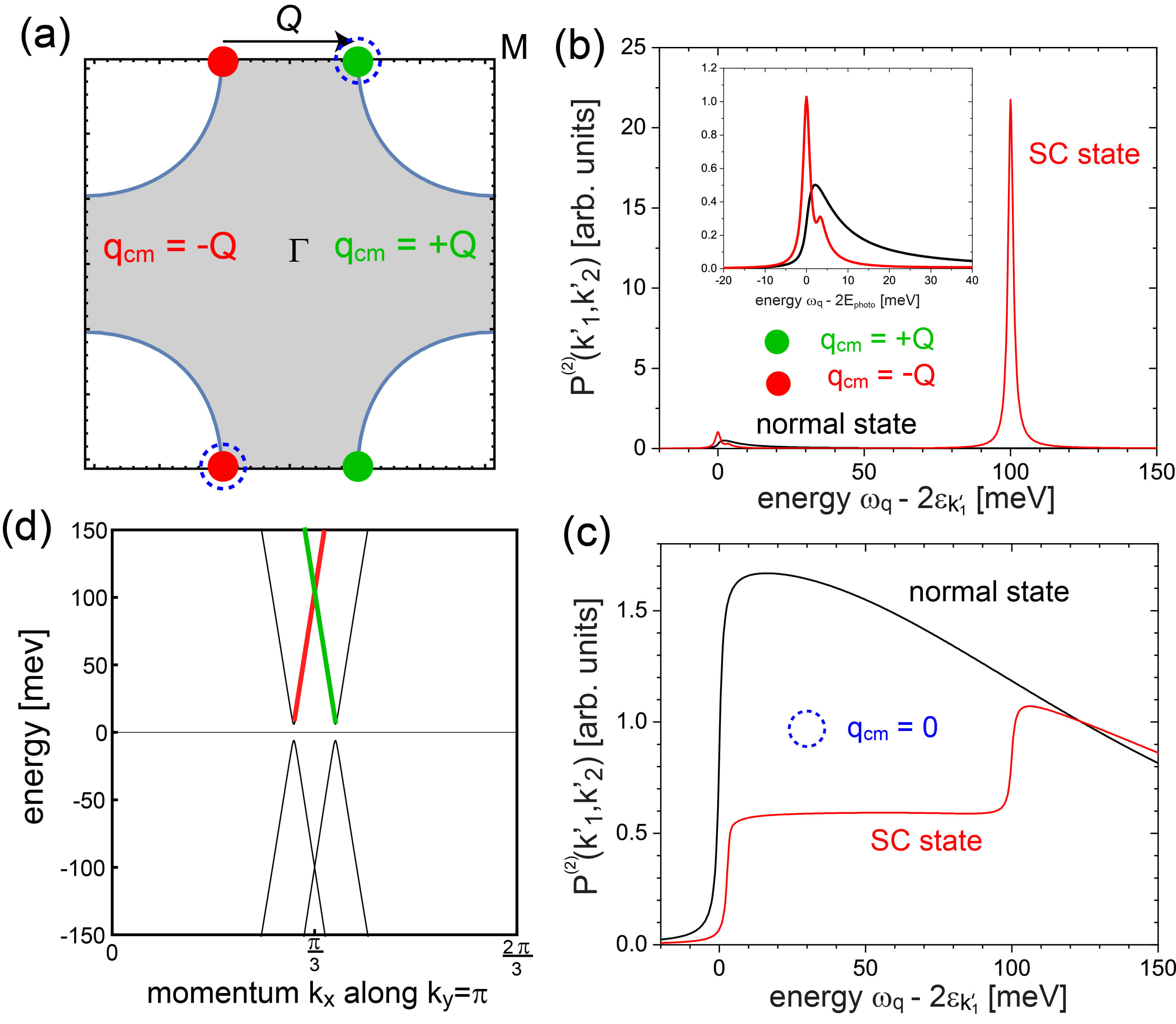}
\caption{(a) Schematic representation of superconducting pairing in the PDW phase, with ${\bf q}_{cm} = \pm {\bf Q}$. For the chosen electronic structure, we have ${\bf Q} = (2\pi/3,0)$. (b) $P^{(2)}$ for photo-electron momenta with ${\bf q}_{cm} = \pm {\bf Q}$, as indicated by the sets of filled red and green circles in (a). The inset shows a zoom-in around $\Delta \omega = 0$. (c) $P^{(2)}$ for photo-electron with opposite momenta, and hence ${\bf q}_{cm}=0$, as indicated by the open blue circles in (a).  (d) Electronic dispersion in the PDW phase as a function of $k_x$ for $k_y = \pi$. }
\label{fig:Fig3}
\end{figure}

To demonstrate the sensitivity of 2$e$-ARPES experiments to detecting the center-of-mass momentum of Cooper pairs, we next consider two distinct superconducting phases with non-zero ${\bf q}_{cm}$. The first is the PDW phase which has been proposed as a possible explanation for the puzzling phenomenology of the pseudo-gap region of the cuprate superconductors \cite{berg2009,lee2020}. In this phase, electronic states with non-zero center-of-mass momentum $+{\bf Q}$ and $-{\bf Q}$ are simultaneously paired, with ${\bf Q}$ connecting the anti-nodal points near $(0,\pm \pi)$, as shown in Fig.~\ref{fig:Fig3}(a).
This leads to a pairing of states near $(\pm {\bf Q}/2 ,\pm \pi)$, such as the ones indicated by red (green) circles in Fig.~\ref{fig:Fig3}(a) with center of mass momentum ${\bf q}_{cm} = \pm {\bf Q}$. For $P^{(2)}$ to directly probe the PDW condensate arising from this pairing, we need to select two photo-electrons with center-of-mass momentum ${\bf q}_{cm} = \pm {\bf Q}$ [red and green circles in Fig.~\ref{fig:Fig3}(a)] as shown in Fig.~\ref{fig:Fig3}(b). $P^{(2)}$ is identical for both sets of photo-electrons, exhibiting a peak at $\Delta \omega = 0$ that is separated by from the continuum by $2 \Delta_{PDW}({{\bf k}^\prime_{1,2}})$. Similar to the case of a uniform $d_{x^2-y^2}$-wave superconductor discussed above, the peak at $\Delta \omega = 0$ directly reflects the existence of a PDW condensate with center of mass momentum ${\bf q}_{cm} = \pm {\bf Q}$. Thus, for photo-electrons with opposite momenta and zero center of mass momentum, as indicated by dashed blue circles in Fig.~\ref{fig:Fig3}(a), $P^{(2)}$ does not exhibit a zero-energy peak, as shown in Fig.~\ref{fig:Fig3}(c). We note that the continuum's peak in Fig.~\ref{fig:Fig3}(b) is considerably higher than was the case for the uniform $d_{x^2-y^2}$-wave case discussed in Fig.~\ref{fig:Fig2}.  The reason for this large intensity is the electronic structure in the PDW phase near ${\bf k}^\prime_{1,2} = ({\bf Q}/2 ,\pm \pi)$, shown in Fig.~\ref{fig:Fig3}(d) where we plot the energy dispersion along $k_x$ for $k_y=\pi$, i.e., perpendicular to the Fermi surface. As before, due to the momentum structure of the Coulomb interaction, the main contribution to $P^{(2)}$ arises from conduction electrons near ${\bf k}_{1,2}^\prime$. The continuum peak arises from the breaking of two Cooper pairs, one of which is located on the red branch of the dispersion, and the other one on the green branch. Due to the linear dispersion near ${\bf k}_{1,2}^\prime$, the energy required to break these two Cooper pairs is essentially constant and equal to $2\Delta_{PDW}$ over an extended range of $k_x$. This implies that, in contrast to the uniform $d_{x^2-y^2}$-wave case, for a fixed $\Delta \omega$ there is an extended momentum range of conduction electron states perpendicular to the Fermi surface that contribute to $P^{(2)}$, yielding the large continuum peak.

\begin{figure}[ht]
\centering
\includegraphics[width=8.5cm]{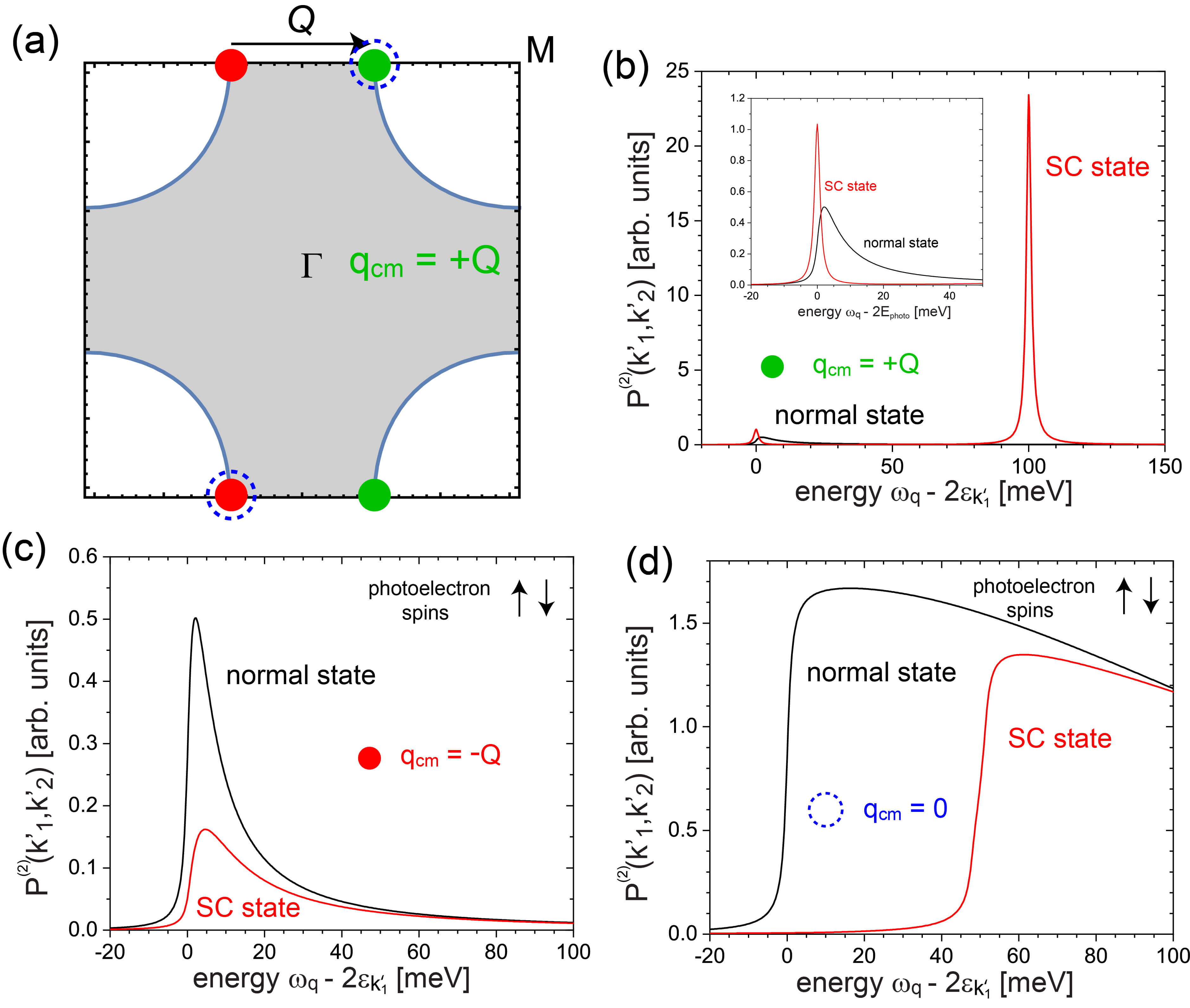}
\caption{(a) Schematic representation of superconducting pairing in the FFLO phase, with ${\bf q}_{cm} = +{\bf Q}$. $P^{(2)}$ for photo-electron momenta with (b) ${\bf q}_{cm} = +{\bf Q}$ [filled green circles in (a), the inset shows a zoom-in around $\Delta \omega = 0$], (c) ${\bf q}_{cm} = -{\bf Q}$ [filled red circles in (a)], and (d) ${\bf q}_{cm}=0$ [open blue circles in (a)].}
\label{fig:Fig4}
\end{figure}

Finally, we consider the FFLO phase where the pairing occurs between states with a single non-zero center-of-mass momentum (strictly speaking, this corresponds to the Fulde-Ferrell phase \cite{Fulde1964}). While there currently is no evidence for an FFLO phase in the cuprate superconductors, the FFLO phase was reported \cite{Bianchi2003} to occur in the heavy fermion $d_{x^2-y^2}$-wave superconductor CeCoIn$_5$ \cite{Allan2013}. To allow explicit comparison with the results for a uniform $d_{x^2-y^2}$-wave superconductor (Fig.~\ref{fig:Fig2}) and the PDW phase
with ${\bf q}_{cm}=\pm {\bf Q}$ (Fig.~\ref{fig:Fig3}), we choose for the FFLO phase ${\bf q}_{cm}=+{\bf Q}$.  By assumption, then, pairing occurs between momentum states with ${\bf k}_1 + {\bf k}_2 = {\bf Q}$, represented by filled green circles in Fig.~\ref{fig:Fig4}(a), but not between states with ${\bf k}_1 + {\bf k}_2 = - {\bf Q}$, as represented by filled red circles in Fig.~\ref{fig:Fig4}(a).
As expected, we find for the FFLO phase that $P^{(2)}$ exhibits a zero-energy peak for ${\bf k}^\prime_1 + {\bf k}^\prime_2 = {\bf Q}$ [green dots in Fig.~\ref{fig:Fig4}(a)] that is separated from the continuum contribution by $2\Delta_{FFLO}$ [Fig.~\ref{fig:Fig4}(b)].
In contrast, momentum states with ${\bf k}^\prime_1 + {\bf k}^\prime_2 = -{\bf Q}$, are unpaired and hence ungapped, such that $P^{(2)}$ in the FFLO phase is simply suppressed in comparison to that in the normal state for these momenta. Furthermore, $P^{(2)}$ for these two momenta does not exhibit a zero-energy peak or a gap towards 2CP continuum excitations [Fig.~\ref{fig:Fig4}(c)], in stark contrast to the PDW phase [Fig.~\ref{fig:Fig3}(b)]. Interestingly, for photo-electrons with opposite momenta [dashed blue circles in Fig.~\ref{fig:Fig4}(a)],  $P^{(2)}$ again exhibits a gap towards 2CP continuum excitations, but its onset energy is shifted from that of the normal state only by $\Delta_{FFLO}$, as only one of the momentum states is paired.

{\it Conclusions}
We have developed a theory for the photo-electron counting rate $P^{(2)}$ measured in 2e-ARPES experiments in a uniform $d_{x^2-y^2}$-wave superconducting, PDW and FFLO phases. A comparison of $P^{(2)}$ shown in Figs.~\ref{fig:Fig2} - \ref{fig:Fig4} demonstrates that 2e-ARPES measurements can identify the center-of-mass momentum (or even multiple center-of-mass momenta, as in the PDW phase), as well as the spin state of Cooper pairs, and thus distinguish between different superconducting pairing states. In addition, it is possible to map out the momentum dependence of the superconducting gap. 2e-ARPES experiments thus provide a valuable new tool for the study of  unconventional superconducting pairing states.

{\it Acknowledgements}
This study was supported by the Center for Quantum Sensing and Quantum Materials, an Energy Frontier Research
Center funded by the U. S. Department of Energy, Office of Science, Basic Energy Sciences under Award DE-SC0021238. P.A. acknowledges support from the EPiQS program of the Gordon and Betty Moore Foundation, Grant No. GBMF9452."

\end{document}


\renewcommand{\thesection}{\arabic{section}}
\renewcommand{\theequation}{S\arabic{equation}}
\renewcommand{\thefigure}{S\arabic{figure}}
\renewcommand{\figurename}{Supplementary Figure}

\title{Distinguishing finite momentum superconducting pairing states with two-electron photoemission spectroscopy\\[0.25cm]
{\large Supplementary Material}}

\author{Fahad Mahmood$^{1,2}$, Thomas Devereaux$^{3,4}$, Peter Abbamonte$^{1,2}$, and Dirk K. Morr$^{5}$}
\affiliation{$^{1}$ Department of Physics, University of Illinois at Urbana-Champaign,Urbana,IL 61801, USA}
\affiliation{$^{2}$ Materials Research Laboratory, University of Illinois at Urbana-Champaign,Urbana,IL 61801, USA}
\affiliation{$^{3}$ Department of Materials Science and Engineering, Stanford University, Stanford, CA, USA}
\affiliation{$^{4}$ Stanford Institute for Materials and Energy Sciences, SLAC National Accelerator Laboratory, Menlo Park, CA, USA}
\affiliation{$^{5}$ University of Illinois at Chicago, Chicago, IL 60607, USA}

\date{\today}
\maketitle

\section{Theoretical Formalism}

As discussed in the main text, for the calculation of the 2e-ARPES photo-electron counting rate, we consider the process shown in Fig.1(a) of the main text. This process is described by the Hamiltonian
\begin{align}
    H_{sc} & = \sum_{{\bf k,q},\sigma,\nu} \gamma_{\nu}({\bf q}) d^\dagger_{{\bf k+q},\sigma} c_{{\bf k},\sigma} \left( a_{{\bf q},\nu}  + a^\dagger_{-{\bf q},\nu}\right)
    + \sum_{{\bf k,p,q},\alpha, \beta} V({\bf q}) d^\dagger_{{\bf k}+{\bf q},\alpha} d^\dagger_{{\bf p}-{\bf q},\beta} d_{{\bf p},\beta} c_{{\bf k},\alpha}  + h.c.
\end{align}
Here, $\gamma_{\nu}({\bf q})$ is the effective electron-photon dipole interaction, $d^\dagger_{{\bf k},\sigma} (c_{{\bf k},\sigma})$ creates (destroys) a photo-electron (conduction electron) with momentum ${\bf k}$ and spin $\sigma$, and $V({\bf q}) \sim [{\bf q}^2 + \kappa^{2}]^{-1}$ is the Fourier transform of the (screened) Coulomb interaction, with $\kappa^{-1}$ being the screening length. For all results shown in the main text, $\kappa^{-1} = 10 a_0$ (see also SM Sec.~\ref{sec:Screening}). Moreover, since the photon momentum is much smaller than typical fermionic momenta, we set it equal to zero, and as the out-of-plane momentum is not conserved upon absorption of the photon,  we take $\gamma_{\nu}({\bf q}) = \gamma_0$ to be independent of the in-plane momentum.

The initial and final states of the entire system, $|\Psi_a \rangle$ and $| \Psi_b \rangle$ respectively, are described by \begin{align}
    | \Psi_a \rangle & = | \Phi_a \rangle | 1_{{\bf q}, \lambda} \rangle_p | 0 \rangle_{pe} \nonumber \\
    | \Psi_b \rangle & = | \Phi_b \rangle | 0 \rangle_p | 1_{{\bf k}_1^\prime, \sigma^\prime_1} 1_{{\bf k}_2^\prime, \sigma^\prime_2} \rangle_{pe} \ .
\end{align}
Here, $|1_{{\bf q} \lambda} \rangle_p$ describes the initial photon state containing one photon with momentum ${\bf q}$ and polarization $\lambda$, and $| 1_{{\bf k}_1^\prime, \sigma^\prime_1} 1_{{\bf k}_2^\prime, \sigma^\prime_2} \rangle_{pe}$ represents the final photo-electron state containing two photo-electrons with momenta ${\bf k}_{1,2}^\prime$ and spin $\sigma^\prime_{1,2}$. The initial and final states of the superconductor are described by $| \Phi_{a,b} \rangle$, respectively. The photo-electron counting rate, $P^{(2)}$, which depends on the two photo-electron momenta and spin projections, is then computed via
\begin{equation}
 P^{(2)}({\bf k}_1^\prime, \sigma^\prime_1, {\bf k}_2^\prime, \sigma^\prime_2 ) =  \frac{1}{Z} \sum_{a,b}  \frac{e^{-\beta E_a}}{\Delta t} \left| \langle  \Psi_b | {\hat S}(\infty,-\infty) | \Psi_a \rangle \right|^2
\end{equation}
where $Z$ is the partition function, the sum runs over all states $|\Phi_{a,b} \rangle$ of the superconductor, and $\Delta t$ is the time over which the photon beam is incident on the superconductor. Expanding the $S$-matrix  ${\hat S}(\infty,-\infty)$ up to second order in $H_{sc}$, and using the above described initial and final states of the system, we obtain
\begin{eqnarray}
&&\left\langle \Psi _b \right\vert \hat{S}%
(\infty ,-\infty )\left\vert \Psi _a
\right\rangle = \nonumber \\
&&-\gamma_{0}\int_{-\Delta t/2}^{\Delta t/2}dt_{2}\int_{t_{2}}^{\infty
}dt_{1}\sum_{{\bf k}\,_{1},{\bf k}_{2}}\delta _{{\bf k}_{1}+{\bf k}_{2}-k_{1}^{\prime
}-{\bf k}_{2}^{\prime },0}\left\{ V({\bf k}_{1}-{\bf k}_{2}^{\prime})\,e^{-i\omega
_{\bf q}t_{2}}e^{i\varepsilon _{{\bf k}_{1}}t_{2}}e^{i\left( -\varepsilon
_{{\bf k}_{1}}+\varepsilon _{{\bf k}_{2}^{\prime }{}}+\varepsilon _{{\bf k}_{1}^{\prime
}}\right) t_{1}}\left\langle \Phi _{b}\right\vert
c_{{\bf k}_{2},\sigma^\prime_2 }(t_{1})c_{{\bf k}_{1},\sigma^\prime_1 }(t_{2})\left\vert \Phi_{a}\right\rangle \right. \nonumber  \\
&& \hspace{1cm} \left. -V({{\bf k}_{1}-{\bf k}_{1}^{\prime }})\,e^{-i\omega _{\bf q}t_{2}}e^{i\varepsilon
_{{\bf k}_{1}}t_{2}}e^{i\left( -\varepsilon _{{\bf k}_{1}}+\varepsilon _{{\bf k}_{1}^{\prime
}}+\varepsilon _{{\bf k}_{2}^{\prime }}\right) t_{1}}\left\langle \Phi _{b} \right\vert c_{{\bf k}_{2},\sigma^\prime_1 }(t_{1})c_{{\bf k}_{1},\sigma^\prime_2
}(t_{2})\left\vert \Phi _{a}\right\rangle \right\}
\label{eq:S}
\end{eqnarray}
with $\omega_{\bf q}$ being the incident photon energy. To further evaluate the above term, we need to rewrite the term involving the fermionic annihilation operators using the respective Bogoliubov transformations for a uniform $d_{x^2-y^2}$-wave superconductor, the PDW and the FFLO phases, which we will consider in the following.

\subsection{2e-ARPES in a uniform $d_{x^2-y^2}$-wave superconductor with zero center-of mass momentum}

\subsubsection{${\bf k}_2^\prime = -{\bf k}_1^\prime$ and $\sigma_2^\prime \not = \sigma_1^\prime$}

For a uniform $d_{x^2-y^2}$-wave superconductor with zero center-of-mass momentum, the Bogoliubov transformation is given by
\begin{eqnarray}
c_{-{\bf k}_{2},\downarrow } &=&-v_{{\bf k}_{2}}\alpha _{{\bf k}_{2}}^{\dagger
}+u_{-{\bf k}_{2}}\beta _{-{\bf k}_{2}} \nonumber \\
c_{{\bf k}_{1},\uparrow } &=&u_{{\bf k}_{1}}\alpha _{{\bf k}_{1}}+v_{-{\bf k}_{1}}\beta
_{-{\bf k}_{1}}^{\dagger }
\end{eqnarray}
At $T=0$, the contribution to $P^{(2)}$ that directly reflects the presence of a superconducting condensate, $P_{SC}^{(2)}$ [see Eq.(4) in the main text], arises from terms of the form
\label{eq:Psc_terms}
\begin{align}
    \left\langle \Phi _{b}\right\vert \alpha_{\bf k}(t_{1}) \alpha^\dagger_{\bf k}(t_{2})\left\vert \Phi_{a} \right\rangle \nonumber \\
    \left\langle \Phi _{b}\right\vert \beta_{\bf k}(t_{1}) \beta^\dagger_{\bf k}(t_{2})\left\vert \Phi_{a}\right\rangle
\end{align}
Terms of these forms, however, can only emerge from Eq.(\ref{eq:S}) if ${\bf k}_2 = - {\bf k}_1$ and $\sigma^\prime_1 \not = \sigma^\prime_2$. Since momentum conservation in Eq.(\ref{eq:S}) requires that
\begin{align*}
   {\bf  k}_{1}+ {\bf  k}_{2}- {\bf  k}_{1}^{\prime}- {\bf  k}_{2}^{\prime} =0
\end{align*}
this immediately implies $ {\bf  k}_{2}^{\prime}=- {\bf  k}_{1}^{\prime}$, i.e., the center-of-mass momentum of the two photo-electrons is zero. Moreover, $\sigma^\prime_1 \not = \sigma^\prime_2$ implies that the two photo-electrons are in a spin-singlet state. Thus, we obtain a non-zero $P_{SC}^{(2)}$ contribution only if the center-of-mass momentum and the spin state of the two photo-electrons is the same as that of the condensate. We then obtain $P^{(2)}=P_{SC}^{(2)}+P_{2cp}^{(2)}$ where
\begin{align}
    P_{SC}^{(2)} & = 2\pi \delta\left(\omega_q-2\varepsilon_{{\bf k}_1^\prime} \right)\frac{1}{Z} \sum_\alpha e^{-\beta E_\alpha} \left\langle \Phi _{a} \right\vert |I({\bf k}_1^\prime)|^2 \left\vert \Phi _{a}\right\rangle
\end{align}
and
\begin{align}
    I({\bf k}_1^\prime) &  = \gamma_0 \sum_{\bf p} V({\bf k}_1^\prime-{\bf p}) \frac{\Delta_{\bf p}}{2E_{\bf p}}\left[ \frac{{\hat n}_{\bf p}^\alpha + {\hat n}_{\bf p}^\beta }{\omega_q + E_{\bf p} - \varepsilon_{\bf p} + i \delta} - \frac{1-{\hat n}_{\bf p}^\alpha + 1- {\hat n}_{\bf p}^\beta }{\omega_q - E_{\bf p} - \varepsilon_{\bf p} + i \delta}\right]
\end{align}
with ${\hat n}_{\bf p}^\alpha = \alpha^\dagger_{\bf p} \alpha_{\bf p}$ and ${\hat n}_{\bf p}^\beta = \beta^\dagger_{\bf p} \beta_{\bf p}$. Moreover, $E_{\bf p} = \sqrt{\xi_{\bf p}^2 + \Delta_{\bf p}^2}$ ($\xi_{\bf p}$) is the electronic dispersion in the superconducting (normal) state. The normal state dispersion is given by
\begin{align}
    \xi_{\bf p} = -2t \left( \cos{p_x}+\cos{p_y}\right) - 4t^\prime \cos{p_x}\cos{p_y} - \mu
\end{align}
with $t^\prime/t=-0.4$, $\mu/t=-0.5$, and $t=300$meV, giving rise to the characteristic cuprate Fermi surface shown in Fig.~2(a) of the main text. Moreover, the superconducting $d_{x^2-y^2}$-wave gap is given by
\begin{align}
    \Delta_{\bf p} = \frac{\Delta_0}{2} \left( \cos{p_x} - \cos{p_y}\right)
\end{align}
with $\Delta_0 = 25$ meV. At $T=0$, $P_{SC}^{(2)}$ simplifies to the expression
which is given in Eq.(4) of the main text. Similarly, we obtain
\begin{align}
    P_{2cp}^{(2)} & = 2\pi \gamma_0^2 \sum_{\bf p} \left[ \delta\left( \omega_q - 2\varepsilon_{{\bf k}_1^\prime}  - 2E_{\bf p}\right) \left\vert \frac{V({\bf k}_1^\prime-{\bf p})}{\omega_q + E_{\bf p} - \varepsilon_{\bf p} + i \delta} \right\vert^2 v_{\bf p}^4 \left\langle \left( 1- {\hat n}_{\bf p}^\alpha \right) \left( 1- {\hat n}_{\bf p}^\beta \right)\right\rangle \right. \nonumber \\
    & \left. + \delta\left( \omega_q - 2\varepsilon_{{\bf k}_1^\prime}   + 2E_{\bf p}\right) \left\vert \frac{V({\bf k}_1^\prime-{\bf p})}{\omega_q - E_{\bf p} - \varepsilon_{\bf p} + i \delta} \right\vert^2 u_{\bf p}^4 \left\langle  {\hat n}_{\bf p}^\alpha {\hat n}_{\bf p}^\beta\right\rangle \right]
    \label{eq:P2}
\end{align}
where $v^2_{\bf p}=\left[1 -\xi_{\bf p}/E_{\bf p}\right]/2$, and $u^2_{\bf p}=\left[1 +\xi_{\bf p}/E_{\bf p}\right]/2$ are the superconducting coherence factors. At $T=0$, this result simplifies to the expression given in Eq.(4) of the main text.\\[0.2cm]

\subsubsection{${\bf k}_2^\prime \not = {\bf k}_1^\prime$ and $\sigma_2^\prime \not = \sigma_1^\prime$}

We next consider the case where the two photo-electrons possess a non-zero center-of-mass momentum, i.e., ${\bf k}_1^\prime + {\bf k}_2^\prime = {\bf l} \not = 0$, and opposite spins. In this case, $P_{SC}^{(2)} \equiv 0$, and $P_{2cp}^{(2)} = P_{\alpha \alpha }^{(2)} + P_{\alpha \beta }^{(2)} + P_{\beta \alpha }^{(2)} + P_{\beta \beta }^{(2)}$ where
\begin{eqnarray}
P_{\alpha \alpha }^{(2)} &=&2\pi \gamma _{0}^{2}\sum_{\mathbf{k}\,_{1},%
\mathbf{k}_{2}}\delta _{\mathbf{k}_{1}+\mathbf{k}_{2},\mathbf{l}}V^{2}(%
\mathbf{k}_{1}-\mathbf{k}_{2}^{\prime })\left\vert \,\frac{1}{\varepsilon _{%
\mathbf{k}_{1}^{\prime }}+\varepsilon _{\mathbf{k}_{2}^{\prime
}}-\varepsilon _{\mathbf{k}_{1}}+E_{\mathbf{k}_{2}}+i\delta }+\frac{1}{%
\varepsilon _{\mathbf{k}_{1}^{\prime }}+\varepsilon _{\mathbf{k}_{2}^{\prime
}}-\varepsilon _{\mathbf{k}_{2}}-E_{\mathbf{k}_{1}}+i\delta }\right\vert
^{2}v_{\mathbf{k}_{2}}^{2}u_{\mathbf{k}_{1}}^{2}\left\langle n_{\mathbf{k}%
_{1}}^{\alpha }\left( 1-n_{\mathbf{k}_{2}}^{\alpha }\right) \right\rangle \nonumber \\
& & \times  \delta \left( \varepsilon _{\mathbf{k}_{1}^{\prime }}+\varepsilon _{\mathbf{k%
}_{2}^{\prime }}-\omega _{0}-E_{\mathbf{k}_{1}}+E_{\mathbf{k}_{2}}\right)  \nonumber \\
P_{\alpha \beta }^{(2)} &=&2\pi \gamma _{0}^{2}\sum_{\mathbf{k}\,_{1},%
\mathbf{k}_{2}}\delta _{\mathbf{k}_{1}+\mathbf{k}_{2},\mathbf{l}}V^{2}(%
\mathbf{k}_{1}-\mathbf{k}_{2}^{\prime })\left\vert \,\frac{1}{\varepsilon _{%
\mathbf{k}_{1}^{\prime }}+\varepsilon _{\mathbf{k}_{2}^{\prime
}}-\varepsilon _{\mathbf{k}_{1}}-E_{\mathbf{k}_{2}}+i\delta }+\,\,\frac{1}{%
\varepsilon _{\mathbf{k}_{1}^{\prime }}+\varepsilon _{\mathbf{k}_{2}^{\prime
}}-\varepsilon _{\mathbf{k}_{2}}-E_{\mathbf{k}_{1}}+i\delta }\right\vert
^{2}u_{\mathbf{k}_{2}}^{2}u_{\mathbf{k}_{1}}^{2}\left\langle n_{\mathbf{k}%
_{1}}^{\alpha }n_{\mathbf{k}_{2}}^{\beta }\right\rangle \nonumber \\
& & \times \delta \left(
\varepsilon _{\mathbf{k}_{1}^{\prime }}+\varepsilon _{\mathbf{k}_{2}^{\prime
}}-\omega _{0}-E_{\mathbf{k}_{1}}-E_{\mathbf{k}_{2}}\right)  \nonumber \\
P_{\beta \alpha }^{(2)} &=&2\pi \gamma _{0}^{2}\sum_{\mathbf{k}\,_{1},%
\mathbf{k}_{2}}\delta _{\mathbf{k}_{1}+\mathbf{k}_{2},\mathbf{l}}V^{2}(%
\mathbf{k}_{1}-\mathbf{k}_{2}^{\prime })\left\vert \,\frac{1}{\varepsilon _{%
\mathbf{k}_{1}^{\prime }}+\varepsilon _{\mathbf{k}_{2}^{\prime
}}-\varepsilon _{\mathbf{k}_{1}}+E_{\mathbf{k}_{2}}+i\delta }+\,\,\frac{1}{%
\varepsilon _{\mathbf{k}_{1}^{\prime }}+\varepsilon _{\mathbf{k}_{2}^{\prime
}}-\varepsilon _{\mathbf{k}_{2}}+E_{\mathbf{k}_{1}}+i\delta }\right\vert
^{2}v_{\mathbf{k}_{2}}^{2}v_{\mathbf{k}_{1}}^{2} \nonumber \\
&&\times \left\langle \left( 1-n_{%
\mathbf{k}_{1}}^{\beta }\right) \left( 1-n_{\mathbf{k}_{2}}^{\alpha }\right)
\right\rangle \delta\left( \varepsilon _{\mathbf{k}_{1}^{\prime
}}+\varepsilon _{\mathbf{k}_{2}^{\prime }}-\omega _{0}+E_{\mathbf{k}_{1}}+E_{%
\mathbf{k}_{2}}\right)  \nonumber \\
P_{\beta \beta }^{(2)} &=&2\pi \gamma _{0}^{2}\sum_{\mathbf{k}\,_{1},\mathbf{%
k}_{2}}\delta _{\mathbf{k}_{1}+\mathbf{k}_{2},\mathbf{l}}V^{2}(\mathbf{k}%
_{1}-\mathbf{k}_{2}^{\prime })\left\vert \,\frac{1}{\varepsilon _{\mathbf{k}%
_{1}^{\prime }}+\varepsilon _{\mathbf{k}_{2}^{\prime }}-\varepsilon _{%
\mathbf{k}_{1}}-E_{\mathbf{k}_{2}}+i\delta }+\frac{1}{\varepsilon _{\mathbf{k%
}_{1}^{\prime }}+\varepsilon _{\mathbf{k}_{2}^{\prime }}-\varepsilon _{%
\mathbf{k}_{2}}+E_{\mathbf{k}_{1}}+i\delta }\right\vert ^{2}v_{\mathbf{k}%
_{1}}^{2}u_{\mathbf{k}_{2}}^{2} \nonumber \nonumber \\
& & \times \left\langle n_{\mathbf{k}_{2}}^{\beta
}\left( 1-n_{\mathbf{k}_{1}}^{\beta }\right) \right\rangle \delta \left(
\varepsilon _{\mathbf{k}_{1}^{\prime }}+\varepsilon _{\mathbf{k}_{2}^{\prime
}}-\omega _{0}+E_{\mathbf{k}_{1}}-E_{\mathbf{k}_{2}}\right)
\end{eqnarray}

\subsubsection{Equal spin polarization of the photo-electrons, and ${\bf k}_2^\prime = - {\bf k}_1^\prime$}

We next consider the case where the two photo-electrons possess the equal spin polarization, and a non-zero center-of-mass momentum, i.e., ${\bf k}_2^\prime = - {\bf k}_1^\prime$. In this case $P^{(2)} = P_{\alpha \alpha }^{(2)} + P_{\alpha \beta }^{(2)} + P_{\beta \beta }^{(2)}$ where
\begin{eqnarray}
P_{\alpha \alpha }^{(2)} &=&4\pi \gamma _{0}^{2}\sum_{\mathbf{k}\,}\left[ V(-%
\mathbf{k}_{1}^{\prime }-\mathbf{k})-V\left( \mathbf{k}_{1}^{\prime }-%
\mathbf{k}\right) \right] ^{2}\left\vert \frac{1}{\omega _{q}-\varepsilon _{%
\mathbf{k}}-E_{\mathbf{k}}+i\delta }\right\vert ^{2}u_{\mathbf{k}%
}^{4}\left\langle n_{\mathbf{k}}^{\alpha }n_{-\mathbf{k}}^{\alpha
}\right\rangle \delta \left( \omega _{q}-2\varepsilon _{\mathbf{k}%
_{1}^{\prime }}+2E_{\mathbf{k}}\right) \nonumber \\
P_{\beta \beta }^{(2)} &=&4\pi \gamma _{0}^{2}\sum_{\mathbf{k}\,}\left[ V(-%
\mathbf{k}_{1}^{\prime }-\mathbf{k})-V\left( \mathbf{k}_{1}^{\prime }-%
\mathbf{k}\right) \right] ^{2}\left\vert \frac{1}{\omega _{q}-\varepsilon _{%
\mathbf{k}}+E_{\mathbf{k}}+\delta }\right\vert ^{2}v_{\mathbf{k}%
}^{4}\left\langle \left( 1-n_{\mathbf{k}}^{\beta }\right) \left( 1-n_{-%
\mathbf{k}}^{\beta }\right) \right\rangle \delta \left( \omega
_{q}-2\varepsilon _{\mathbf{k}_{1}^{\prime }}-2E_{\mathbf{k}}\right) \nonumber \\
P_{\alpha \beta }^{(2)} &=&2\pi \gamma _{0}^{2}\delta \left( \omega
_{q}-2\varepsilon _{\mathbf{k}_{1}^{\prime }}\right) \sum_{\mathbf{k}\,}\,%
\left[ V(-\mathbf{k}_{1}^{\prime }-\mathbf{k})-V\left( \mathbf{k}%
_{1}^{\prime }-\mathbf{k}\right) \right] ^{2}\left\vert \frac{1}{\omega
_{q}-\varepsilon _{\mathbf{k}}-E_{\mathbf{k}}+i\delta }+\,\frac{1}{\omega
_{q}-\varepsilon _{\mathbf{k}}+E_{\mathbf{k}}+i\delta }\right\vert
^{2} \nonumber \\
& & \times \left( \frac{\Delta _{k}}{2E_{k}}\right) ^{2}\left\langle \left( 1-n_{%
\mathbf{k}}^{\beta }\right) n_{\mathbf{k}}^{\alpha }\right\rangle
\end{eqnarray}
Note that for $T \not = 0$, $P_{\alpha \beta }^{(2)}$ can also exhibit a peak for $\Delta \omega = \omega_{q}-2\varepsilon_{\bf{k}_{1}^{\prime} }= 0$,  similar to $P_{SC}^{(2)}$ in Eq.(4) of the main text. However, the term $\left\langle \left( 1-n_{\mathbf{k}}^{\beta }\right) n_{\mathbf{k}}^{\alpha }\right\rangle$ in $P_{\alpha \beta }^{(2)}$ implies that it vanishes for $T=0$, as two electrons with equal spin can only be ejected from the sample at $\Delta \omega = 0$ if Cooper pairs have been broken by thermal excitations.

\subsection{2e-ARPES in the FFLO phase}

We next consider the photo-electron counting rate in the FFLO phase \cite{Fulde64,Larkin65}, where superconducting pairing occurs between two electrons with center-of-mass momentum ${\bf Q}$ (strictly speaking, this corresponds to the Fulde-Ferrell phase \cite{Fulde64}), as described by the mean-field Hamiltonian
\begin{eqnarray}
H &=&\sum_{{\bf k},\sigma }\xi _{{\bf k}}c_{{\bf k},\sigma }^{\dagger }c_{{\bf k},\sigma
}-\sum_{{\bf k}} \left(\Delta_{FF}({\bf k})c_{{\bf k}+{\bf Q}/2,\uparrow }^{\dagger }c_{-{\bf k}+{\bf Q}/2,\downarrow
}^{\dagger }+\Delta_{FF}({\bf k})c_{-{\bf k}+{\bf Q}/2,\downarrow }c_{{\bf k}+{\bf Q}/2,\uparrow } \right)
\end{eqnarray}
Note that $\Delta_{FF}({\bf k})$ also depends on ${\bf Q}$, and in general needs to be self-consistently computed \cite{Gra19}. However, since in the main text, we consider only photo-electron momenta near the antinodal points (see Fig.~4 in the main text), and due the suppression of large momentum transfer due to the Coulomb interaction, we can neglect the detailed momentum dependence of the superconducting gap in the FF phase and simply set $\Delta_{FF}$ equal to a constant value, with $\Delta_{FF} = 50$ meV.

To diagonalize the Hamiltonian, we next use the Bogoliubov transformation%
\begin{eqnarray}
c_{{\bf k}+{\bf Q}/2,\uparrow }^{\dagger } &=&u_{{\bf k}}\alpha _{{\bf k}+{\bf Q}/2}^{\dagger }+v_{{\bf k}}\beta
_{-{\bf k}+{\bf Q}/2} \nonumber \\
c_{-{\bf k}+{\bf Q}/2,\downarrow } &=&-v_{{\bf k}}\alpha _{{\bf k}+{\bf Q}/2}^{\dagger }+v_{{\bf k}}\beta
_{-{\bf k}+{\bf Q}/2}
\end{eqnarray}%
which yields
\begin{eqnarray}
H &=&\sum_{{\bf k}}E_{{\bf k}+{\bf Q}/2}^{\alpha }\alpha _{{\bf k}+{\bf Q}/2}^{\dagger }\alpha
_{{\bf k}+{\bf Q}/2}+E_{-{\bf k}+{\bf Q}/2}^{\beta }\beta _{-{\bf k}+{\bf Q}/2}^{\dagger }\beta _{-{\bf k}+{\bf Q}/2}
\end{eqnarray}%
where
\begin{align}
    E_{\bf k+Q/2}^\alpha &= \sqrt{ \left( \varepsilon_{\bf k}^{+}\right)^2 + \Delta^2_{FF}({\bf k}) } + \varepsilon_{\bf k}^{-} \nonumber \\
    E_{\bf -k+Q/2}^\beta &= \sqrt{ \left( \varepsilon_{\bf k}^{+}\right)^2 + \Delta^2_{FF}({\bf k})  } - \varepsilon_{\bf k}^{-}  \nonumber \\
      \varepsilon_{\bf k}^{\pm} &= \frac{\varepsilon_{\bf k + Q/2} \pm \varepsilon_{\bf -k + Q/2}}{2}
\end{align}
and the coherence factor are given by
\begin{align}
      u_{\bf k} v_{\bf k} &= \frac{\Delta_{FF}({\bf k}) }{2\sqrt{ \left( \varepsilon_{\bf k}^{+}\right)^2 + \Delta^2_{FF}({\bf k})  }} \nonumber \\
      u_{\bf k}^2 &= \frac{1}{2} \left[ 1+ \frac{\varepsilon_{\bf k}^{+}}{\sqrt{ \left( \varepsilon_{\bf k}^{+}\right)^2 + \Delta^2_{FF}({\bf k})  }}\right] \nonumber \\
      v_{\bf k}^2 &= \frac{1}{2} \left[ 1 - \frac{\varepsilon_{\bf k}^{+}}{\sqrt{ \left( \varepsilon_{\bf k}^{+}\right)^2 + \Delta^2_{FF}({\bf k})  }}\right] \nonumber \\
\end{align}

\subsubsection{${\bf k}_1^\prime + {\bf k}_2^\prime = {\bf Q}$}

In order to obtain a non-zero $P_{SC}^{(2)}$ in the FF phase,  Eq.(\ref{eq:S}) needs to contain terms of the form
\begin{align}
    \left\langle \Phi _{b}\right\vert \alpha_{\bf k +Q/2}(t_{1}) \alpha^\dagger_{\bf k+Q/2}(t_{2})\left\vert \Phi_{a} \right\rangle \nonumber \\
    \left\langle \Phi _{b}\right\vert \beta_{\bf -k+Q/2}(t_{1}) \beta^\dagger_{\bf -k+Q/2}(t_{2})\left\vert \Phi_{a}\right\rangle
\end{align}
which together with momentum conservation
\begin{align*}
   {\bf k}+{\bf Q}/2 + \left( -{\bf k}+{\bf Q}/2 \right)-{\bf k}_{1}^{\prime}-{\bf k}_{2}^{\prime} =0
\end{align*}
implies  ${\bf k}_{1}^{\prime}+{\bf k}_{2}^{\prime} = {\bf Q}$, requiring that the center-of-mass momentum of the two photo-electrons be ${\bf Q}$. In this case, we then obtain $P^{(2)}=P_{SC}^{(2)}+P_{2cp}^{(2)}$ where
\begin{align}
    P_{SC}^{(2)} & = 2\pi \delta\left(\omega_{\bf q} - \varepsilon_{{\bf k}_1^\prime} - \varepsilon_{{\bf k}_2^\prime}  \right)\frac{1}{Z} \sum_\alpha e^{-\beta E_\alpha} \left\langle \Psi _{s}^{\alpha} \right\vert |I({\bf k}_1^\prime)|^2 \left\vert \Psi _{s}^{\alpha }\right\rangle
\end{align}
and
\begin{align}
I({\bf k}_{1}^{\prime }) &= \gamma_0\sum_{\bf p} u_{\bf p}v_{\bf p} V({\bf k}_1^\prime-(-{\bf p}+{\bf Q}/2))\left[ \frac{1-n_{-\bf p+Q/2}^{\beta }}{\omega_{\bf q}-\varepsilon_{\bf p+Q/2}-E_{-\bf p+Q/2}^{\beta }-i\delta } +\frac{1-n_{\bf p+Q/2}^{\alpha }}{\omega_{\bf q}-\varepsilon_{-\bf p+Q/2}-E_{\bf p+Q/2}^{\alpha }-i\delta } \right. \nonumber  \\
& \left. -\frac{n_{\bf p+Q/2}^{\alpha }}{\omega_{\bf q}-\varepsilon_{\bf p +Q/2}+E_{\bf p +Q/2}^{\alpha }-i\delta }-
\frac{n_{-\bf p +Q/2}^{\beta }}{\omega_{\bf q}-\varepsilon _{-\bf p+Q/2}+E_{-\bf p +Q/2}^{\beta }-i\delta }%
\right]
\end{align}
Similarly, we obtain
\begin{eqnarray}
P_{2cp}^{(2)} &=&2\pi \gamma_0^2\sum_{\bf p} V^2({\bf k}_1^\prime-(-{\bf p}+{\bf Q}/2)) \left\{ \left\vert \frac{1}{
-\varepsilon _{\bf p+Q/2}+\varepsilon _{{\bf k}_{2}^{\prime }{}}+\varepsilon
_{{\bf k}_{1}^{\prime }}+E_{\bf p+Q/2}^{\alpha }+i\delta }+\,\frac{1}{-\varepsilon
_{-\bf p+Q/2}+\varepsilon _{{\bf k}_{1}^{\prime }}+\varepsilon _{{\bf k}_{2}^{\prime
}}+E_{-\bf p+Q/2}^{\beta }+i\delta }\right\vert^{2} \right. \nonumber \\
&\times  & v_{\bf p}^{4}\left\langle \left(
1-n_{-\bf p+Q/2}^{\beta }\right) \left( 1-n_{\bf p+Q/2}^{\alpha }\right)
\right\rangle \delta \left( \varepsilon _{{\bf k}_{2}^{\prime }}+\varepsilon
_{{\bf k}_{1}^{\prime }}-\omega _{0}+E_{-\bf p+Q/2}^{\beta }+E_{\bf p+Q/2}^{\alpha
}\right) \nonumber \\
&+& \left\vert \frac{1}{-\varepsilon
_{\bf p+Q/2}+\varepsilon _{{\bf k}_{2}^{\prime }{}}+\varepsilon _{{\bf k}_{1}^{\prime
}}-E_{-\bf p+Q/2}^{\beta }+i\delta }+\frac{1}{-\varepsilon
_{-\bf p+Q/2}+\varepsilon _{{\bf k}_{1}^{\prime }}+\varepsilon _{{\bf k}_{2}^{\prime
}}-E_{\bf p+Q/2}^{\alpha }+i\delta }\right\vert^{2} \nonumber \\
&\times& \left.  u_{\bf p}^{4}\left\langle
n_{-\bf p+Q/2}^{\beta }n_{\bf p+Q/2}^{\alpha }\right\rangle \delta \left( \varepsilon
_{{\bf k}_{2}^{\prime }}+\varepsilon _{{\bf k}_{1}^{\prime }}-\omega
_{0}-E_{\bf p+Q/2}^{\alpha }-E_{-\bf p+Q/2}^{\beta }\right) \right\}
\end{eqnarray}

\subsubsection{${\bf k}_1^\prime + {\bf k}_2^\prime \not = {\bf Q}$}

We next consider the case when ${\bf k}_1^\prime + {\bf k}_2^\prime = {\bf l} \not = {\bf Q}$
In this case, we obtain $P^{(2)} = P_{\alpha\alpha}^{(2)} + P_{\alpha\beta}^{(2)} + P_{\beta\alpha}^{(2)} + P_{\beta\beta}^{(2)}$
where
\begin{eqnarray}
P_{\alpha \alpha }^{(2)} &=&2\pi \gamma _{0}^{2}\sum_{\mathbf{k}\,_{1},%
\mathbf{k}_{2}}\delta _{\mathbf{k}_{2},\mathbf{k}_{1}-\mathbf{l}+\mathbf{Q}%
\;}\,\left\vert
\frac{1}{-\varepsilon _{\mathbf{k}_{1}+\mathbf{Q}/2}+\varepsilon _{\mathbf{k}%
_{2}^{\prime }{}}+\varepsilon _{\mathbf{k}_{1}^{\prime }}+E_{\mathbf{k}_{2}+%
\mathbf{Q}/2}^{\alpha }+i\delta } +\frac{1}{-\varepsilon _{\mathbf{k}_{2}+%
\mathbf{Q}/2}+\varepsilon _{\mathbf{k}_{2}^{\prime }{}}+\varepsilon _{%
\mathbf{k}_{1}^{\prime }}-E_{\mathbf{k}_{1}+\mathbf{Q}/2}^{\alpha }+i\delta }%
\right\vert^{2} \nonumber \\
&&\times V^{2}(\mathbf{k}_{2}^{\prime }-\mathbf{k}_{1}-\mathbf{Q}/2) \left( v_{\mathbf{k}_{2}}u_{\mathbf{k}_{1}}\right) ^{2}\left\langle
n_{\mathbf{k}_{1}+\mathbf{Q}/2}^{\alpha }\left( 1-n_{\mathbf{k}_{2}+\mathbf{Q%
}/2}^{\alpha }\right) \right\rangle \delta \left( \varepsilon _{\mathbf{k}%
_{2}^{\prime }}+\varepsilon _{\mathbf{k}_{1}^{\prime }}-\omega _{0}+E_{%
\mathbf{k}_{2}+\mathbf{Q}/2}^{\alpha }-E_{\mathbf{k}_{1}+\mathbf{Q}%
/2}^{\alpha }\right) \nonumber \\
P_{\alpha \beta }^{(2)} &=&2\pi \gamma _{0}^{2}\sum_{\mathbf{k}\,_{1},%
\mathbf{k}_{2}} \delta _{\mathbf{k}_{2},\mathbf{k}_{1}-\mathbf{l}+\mathbf{Q}%
\;} \left\vert
\frac{1}{-\varepsilon _{\mathbf{k}_{1}+\mathbf{Q}/2}+\varepsilon _{\mathbf{k}%
_{2}^{\prime }{}}+\varepsilon _{\mathbf{k}_{1}^{\prime }}+E_{\mathbf{k}_{2}+%
\mathbf{Q}/2}^{\alpha }+i\delta }+\frac{1}{-\varepsilon _{\mathbf{k}_{2}+%
\mathbf{Q}/2}+\varepsilon _{\mathbf{k}_{2}^{\prime }{}}+\varepsilon _{%
\mathbf{k}_{1}^{\prime }}+E_{-\mathbf{k}_{1}+\mathbf{Q}/2}^{\beta }+i\delta }%
\right\vert ^{2} \nonumber \\
&&\times V^{2}(\mathbf{k}_{2}^{\prime }-\mathbf{k}_{1}-\mathbf{Q}/2)\, \left( v_{\mathbf{k}_{2}}v_{\mathbf{k}_{1}}\right) ^{2}\left\langle
\left( 1-n_{-\mathbf{k}_{1}+\mathbf{Q}/2}^{\beta }\right) \left( 1-n_{%
\mathbf{k}_{2}+\mathbf{Q}/2}^{\alpha }\right) \right\rangle \delta \left(
\varepsilon _{\mathbf{k}_{2}^{\prime }}+\varepsilon _{\mathbf{k}_{1}^{\prime
}}-\omega _{0}+E_{\mathbf{k}_{2}+\mathbf{Q}/2}^{\alpha }+E_{-\mathbf{k}_{1}+%
\mathbf{Q}/2}^{\beta }\right) \nonumber \\
P_{\beta \alpha }^{(2)} &=&2\pi \gamma _{0}^{2}\sum_{\mathbf{k}\,_{1},%
\mathbf{k}_{2}}\delta _{\mathbf{k}_{2},\mathbf{k}_{1}-\mathbf{l}+\mathbf{Q}%
\;}\left\vert
\frac{1}{-\varepsilon _{\mathbf{k}_{1}+\mathbf{Q}/2}+\varepsilon _{\mathbf{k}%
_{2}^{\prime }{}}+\varepsilon _{\mathbf{k}_{1}^{\prime }}-E_{-\mathbf{k}_{2}+%
\mathbf{Q}/2}^{\beta }+i\delta }+\frac{1}{-\varepsilon _{\mathbf{k}_{2}+%
\mathbf{Q}/2}+\varepsilon _{\mathbf{k}_{2}^{\prime }{}}+\varepsilon _{%
\mathbf{k}_{1}^{\prime }}-E_{\mathbf{k}_{1}+\mathbf{Q}/2}^{\alpha }+i\delta }%
\right\vert ^{2} \nonumber \\
&&\times V^{2}(\mathbf{k}_{2}^{\prime }-\mathbf{k}_{1}-\mathbf{Q}/2)\,\left( u_{\mathbf{k}_{2}}u_{\mathbf{k}_{1}}\right) ^{2}\left\langle
n_{\mathbf{k}_{1}+\mathbf{Q}/2}^{\alpha }n_{-\mathbf{k}_{2}+\mathbf{Q}%
/2}^{\beta }\right\rangle \delta \left( \varepsilon _{\mathbf{k}_{2}^{\prime
}}+\varepsilon _{\mathbf{k}_{1}^{\prime }}-\omega _{0}-E_{\mathbf{k}_{1}+%
\mathbf{Q}/2}^{\alpha }-E_{-\mathbf{k}_{2}+\mathbf{Q}/2}^{\beta }\right) \nonumber \\
P_{\beta \beta }^{(2)} &=&2\pi \gamma _{0}^{2}\sum_{\mathbf{k}\,_{1},\mathbf{%
k}_{2}}\delta _{\mathbf{k}_{2},\mathbf{k}_{1}-\mathbf{l}+\mathbf{Q}\;}\left\vert \frac{1}{%
-\varepsilon _{\mathbf{k}_{1}+\mathbf{Q}/2}+\varepsilon _{\mathbf{k}%
_{2}^{\prime }{}}+\varepsilon _{\mathbf{k}_{1}^{\prime }}-E_{-\mathbf{k}_{2}+%
\mathbf{Q}/2}^{\beta }+i\delta }+\frac{1}{-\varepsilon _{\mathbf{k}_{2}+%
\mathbf{Q}/2}+\varepsilon _{\mathbf{k}_{2}^{\prime }{}}+\varepsilon _{%
\mathbf{k}_{1}^{\prime }}+E_{-\mathbf{k}_{1}+\mathbf{Q}/2}^{\beta }+i\delta }%
\right\vert ^{2} \nonumber \\
&&\times V^{2}(%
\mathbf{k}_{2}^{\prime }-\mathbf{k}_{1}-\mathbf{Q}/2)\, \left( u_{\mathbf{k}_{2}}v_{\mathbf{k}_{1}}\right) ^{2}\left\langle
\left( 1-n_{-\mathbf{k}_{1}+\mathbf{Q}/2}^{\beta }\right) n_{-\mathbf{k}_{2}+%
\mathbf{Q}/2}^{\beta }\right\rangle \delta \left( \varepsilon _{\mathbf{k}%
_{2}^{\prime }}+\varepsilon _{\mathbf{k}_{1}^{\prime }}-\omega _{0}+E_{-%
\mathbf{k}_{1}+\mathbf{Q}/2}^{\beta }-E_{-\mathbf{k}_{2}+\mathbf{Q}%
/2}^{\beta }\right) \nonumber \\
\end{eqnarray}

\subsection{2e-ARPES in the PDW phase}

We next consider the 2e-ARPES photo-electron counting rate $P^{(2)}$ in the PDW phase, whose mean-field Hamiltonian is given by
\begin{eqnarray}
H &=&{\sum_{\mathbf{k}}}^{\prime }\left( c_{\mathbf{k},\uparrow }^{\dagger
},c_{\mathbf{k}+\mathbf{Q},\uparrow }^{\dagger },c_{\mathbf{k}-\mathbf{Q}%
,\uparrow }^{\dagger },c_{-\mathbf{k},\downarrow },c_{-\mathbf{k}+\mathbf{Q}%
,\downarrow },c_{-\mathbf{k}-\mathbf{Q},\downarrow }\right) \left(
\begin{array}{cccccc}
\xi _{\mathbf{k}} & 0 & 0 & 0 & \Delta  & \Delta  \\
0 & \xi _{\mathbf{k}+\mathbf{Q}} & 0 & \Delta  & \Delta  & 0 \\
0 & 0 & \xi _{\mathbf{k}-\mathbf{Q}} & \Delta  & 0 & \Delta  \\
0 & \Delta  & \Delta  & -\xi _{\mathbf{k}} & 0 & 0 \\
\Delta  & \Delta  & 0 & 0 & -\xi _{\mathbf{k}-\mathbf{Q}} & 0 \\
\Delta  & 0 & \Delta  & 0 & 0 & -\xi _{\mathbf{k}+\mathbf{Q}}%
\end{array}%
\right) \left(
\begin{array}{c}
c_{\mathbf{k},\uparrow } \\
c_{\mathbf{k}+\mathbf{Q},\uparrow } \\
c_{\mathbf{k}-\mathbf{Q},\uparrow } \\
c_{-\mathbf{k},\downarrow }^{\dagger } \\
c_{-\mathbf{k}+\mathbf{Q},\downarrow }^{\dagger } \\
c_{-\mathbf{k}-\mathbf{Q},\downarrow }^{\dagger }%
\end{array}%
\right) \nonumber \\
&=&{\sum_{\mathbf{k}}}^{\prime }\Psi _{\mathbf{k}}^{\dagger }\hat{H}_{\mathbf{k%
}}\Psi _{\mathbf{k}}
\end{eqnarray}%
where the primed sum runs over the Brillouin zone of the PDW phase. Here, $\Delta = \Delta_{PDW}$ is the superconducting gap in the PDW phase which depends on ${\bf k}$ and ${\bf Q}$, and in general needs to be self-consistently computed. However, since in the main text, we consider only photo-electron momenta near the antinodal points (see Fig.~3 in the main text), and due the suppression of large momentum transfer due to the Coulomb interaction, we can neglect the detailed momentum dependence of the superconducting gap in the PDW phase and simply set $\Delta_{PDW}$ equal to a constant value, with $\Delta_{PDW} = 50$ meV.

We next diagonalize the Hamiltonian using the unitary transformation%
\begin{eqnarray}
\left( c_{\mathbf{k},\uparrow }^{\dagger },c_{\mathbf{k}+Q,\uparrow
}^{\dagger },c_{\mathbf{k}-Q,\uparrow }^{\dagger },c_{-\mathbf{k},\downarrow
},c_{-\mathbf{k}+Q,\downarrow },c_{-\mathbf{k}-Q,\downarrow }\right)
&=&\left( \gamma _{1,\mathbf{k}}^{\dagger },\gamma _{2,\mathbf{k}}^{\dagger
},\gamma _{3\mathbf{k}}^{\dagger },\gamma _{4,\mathbf{k},}^{\dagger },\gamma
_{5,\mathbf{k}}^{\dagger },\gamma _{6,\mathbf{k}}^{\dagger }\right) \hat{U}_{%
\mathbf{k}}^{\dagger }=\Gamma _{\mathbf{k}}^{\dagger }\hat{U}_{\mathbf{k}%
}^{\dagger } \nonumber \\
\left(
\begin{array}{c}
c_{\mathbf{k},\uparrow } \\
c_{\mathbf{k}+Q,\uparrow } \\
c_{\mathbf{k}-Q,\uparrow } \\
c_{-\mathbf{k},\downarrow }^{\dagger } \\
c_{-\mathbf{k}+Q,\downarrow }^{\dagger } \\
c_{-\mathbf{k}-Q,\downarrow }^{\dagger }%
\end{array}%
\right)  &=&\hat{U}\left(
\begin{array}{c}
\gamma _{1,\mathbf{k}} \\
\gamma _{2,\mathbf{k}} \\
\gamma _{3\mathbf{k}} \\
\gamma _{4,\mathbf{k},} \\
\gamma _{5,\mathbf{k}} \\
\gamma _{6,\mathbf{k}}%
\end{array}%
\right) =\hat{U}_{\mathbf{k}}\Gamma _{\mathbf{k}}
\end{eqnarray}%
with $\hat{U}$ being a unitary matrix consisting of the eigenvectors of $%
\hat{H}_{\mathbf{k}}$. Thus, we obtain%
\begin{align}
H=&{\sum_{\mathbf{k}}}^{\prime }\Psi _{\mathbf{k}}^{\dagger }\hat{H}_{\mathbf{k}%
}\Psi _{\mathbf{k}}={\sum_{\mathbf{k}}}^{\prime }\Gamma _{\mathbf{k}}^{\dagger
}\hat{U}_{\mathbf{k}}^{\dagger }\hat{H}_{\mathbf{k}}\hat{U}_{\mathbf{k}%
}\Gamma _{\mathbf{k}}={\sum_{\mathbf{k}}}^{\prime }\Gamma _{\mathbf{k}%
}^{\dagger }\hat{E}_{\mathbf{k}}\Gamma _{\mathbf{k}}
\end{align}
with%
\begin{align}
\hat{E}_{\mathbf{k}}=\left(
\begin{array}{cccccc}
E_{1,\mathbf{k}} & 0 & 0 & 0 & 0 & 0 \\
0 & E_{2,\mathbf{k}} & 0 & 0 & 0 & 0 \\
0 & 0 & E_{3,\mathbf{k}} & 0 & 0 & 0 \\
0 & 0 & 0 & E_{4,\mathbf{k}} & 0 & 0 \\
0 & 0 & 0 & 0 & E_{5,\mathbf{k}} & 0 \\
0 & 0 & 0 & 0 & 0 & E_{6,\mathbf{k}}%
\end{array}%
\right)
\end{align}
and $E_{i,\mathbf{k}}(i=1,..,6)$ are the eigenenergies of $\hat{H}_{\mathbf{k}}$.

\subsubsection{${\bf k}_1^\prime + {\bf k}_2^\prime = \pm{\bf Q}$}

In this case, we obtain $P^{(2)}=P_{SC}^{(2)}+P_{2cp}^{(2)} $
where
\begin{align}
P_{SC}^{(2)}=&2\pi \delta \left( \omega _{\bf q}-\varepsilon _{k_{2}^{\prime }}-\varepsilon
_{k_{1}^{\prime }}\right) \frac{1}{Z}\sum_{\alpha \,\beta
}e^{-\beta E_{\alpha }}\left\langle \Psi _{s}^{\alpha }\right\vert
\left\vert I(\mathbf{k}_{1}^{\prime })\right\vert
^{2}\left\vert \Psi _{s}^{\alpha }\right\rangle
\end{align}
with
\begin{align}
I(\mathbf{k}_{1}^{\prime })=\gamma _{0}\sum_{%
\mathbf{p}}V\left( \mathbf{k}_{1}^{\prime }-\left( \mathbf{p}+\mathbf{Q}%
\right) \right) \,\sum_{i=1,6}\left[ {\hat U}_{\mathbf{p}}\right] _{5i}\left[ {\hat U}_{%
\mathbf{p}}\right] _{1i}\left[ \frac{n_{F}^{(i)}(\mathbf{p})}{\omega _{\bf q}-\varepsilon _{\mathbf{p}}+E_{i,\mathbf{p}}+i\delta }-\frac{1-n_{F}^{(i)}(%
\mathbf{p})}{\omega _{\bf q}-\varepsilon _{\mathbf{p}}-E_{i,\mathbf{p}}+i\delta }\right]
\end{align}
with $\left[ {\hat U}_{\mathbf{p}}\right] _{ij}$ being the $(ij)$ element of the matrix ${\hat U}_{\mathbf{p}}$, and
\begin{eqnarray}
P_{2cp}^{(2)} &=&2\pi \gamma _{0}^{2}\sum_{\mathbf{p}\,}V^{2}\left( \mathbf{k%
}_{1}^{\prime }-(-\mathbf{p}+\mathbf{Q})\right) \sum_{i\neq j}\left( \left[
{\hat U}_{\mathbf{p}}\right] _{5i}\left[ {\hat U}_{\mathbf{p}}\right] _{1j}\right)
^{2}\left\vert \frac{1}{-\varepsilon _{\mathbf{p}}+\varepsilon _{\mathbf{k}%
_{2}^{\prime }}+\varepsilon _{\mathbf{k}_{1}^{\prime }}+E_{i,\mathbf{p}%
}+i\delta }+\frac{1}{-\varepsilon _{\mathbf{p}}+\varepsilon _{\mathbf{k}%
_{2}^{\prime }}+\varepsilon _{\mathbf{k}_{1}^{\prime }}-E_{j,\mathbf{p}%
}+i\delta }\right\vert ^{2} \nonumber \\
&&\delta \left( \varepsilon _{\mathbf{k}_{2}^{\prime }}+\varepsilon _{%
\mathbf{k}_{1}^{\prime }}-\omega _{0}+E_{i,\mathbf{p}}-E_{j,\mathbf{p}%
}\right) \left\langle \left( 1-n_{F}^{(i)}(\mathbf{p})\right) n_{F}^{(j)}(%
\mathbf{p})\right\rangle
\end{eqnarray}

\subsubsection{${\bf k}_1^\prime + {\bf k}_2^\prime = 0$}

In this case, we obtain

\begin{align}
    P^{(2)}&=2\pi \gamma _{0}^{2}\sum_{\mathbf{p}\,}V^{2}\left( \mathbf{k}%
_{1}^{\prime }-\mathbf{p}\right) \,\sum_{i\neq j}\left( \left[ {\hat U}_{\mathbf{p}}%
\right] _{4i}\left[ {\hat U}_{\mathbf{p}}\right] _{1j}\right) ^{2}\left\vert \frac{1%
}{\varepsilon _{\mathbf{k}_{1}^{\prime }}+\varepsilon _{\mathbf{k}%
_{2}^{\prime }}-\varepsilon _{\mathbf{p}}+E_{i,\mathbf{p}}+i\delta }+\frac{1%
}{\varepsilon _{\mathbf{k}_{1}^{\prime }}+\varepsilon _{\mathbf{k}%
_{2}^{\prime }}-\varepsilon _{\mathbf{p}}-E_{j,\mathbf{p}_{1}}+i\delta }%
\right\vert ^{2} \nonumber \\
& \times \left\langle n_{F}^{(j)}(\mathbf{p})\left( 1-n_{F}^{(i)}(%
\mathbf{p})\right) \right\rangle \delta \left( \varepsilon _{\mathbf{k}%
_{1}^{\prime }}+\varepsilon _{\mathbf{k}_{2}^{\prime }}-\omega _{0}+E_{i,\mathbf{p}}-E_{j,\mathbf{p}}\right)
\end{align}

\section{Auger process contribution to the $2e$-ARPES photo-electron counting rate for a uniform $d_{x^2-y^2}$-wave superconductor}

We next consider the contribution to the 2e-ARPES photo-electron counting rate involving the Auger process, $P_{Aug}^{(2)}$, shown in Fig.1(b) of the main text. In a uniform $d_{x^2-y^2}$-wave superconductor, for two photo-electrons with momenta ${\bf k}_2^\prime = -{\bf k}_1^\prime$ and opposite spin, $\sigma_2^\prime \not = \sigma_1^\prime$, we obtain at $T=0$
\begin{eqnarray}
P_{Aug}^{(2)} &=&8\pi \delta \left( 2\varepsilon _{\mathbf{k}_{1}^{\prime
}}-\omega _{0}\right) \left\langle n_{\mathbf{k}_{1}^{\prime
}}^{f}\right\rangle \gamma _{0}^{2}\left\vert \frac{1}{\varepsilon _{\mathbf{%
k}_{1}^{\prime }}+\zeta _{\mathbf{k}_{1}^{\prime }}+i\delta }\right\vert
^{2}\left\vert \sum_{\mathbf{p}}U\left( \mathbf{p}-\mathbf{k}_{1}^{\prime
}\right) \frac{\Delta _{\mathbf{p}}}{2E_{\mathbf{p}}}\right\vert ^{2} \nonumber \\
&&+8\pi \gamma _{0}^{2}\left\langle n_{\mathbf{k}_{1}^{\prime
}}^{f}\right\rangle \,\sum_{\mathbf{p}}\delta \left( 2\varepsilon _{\mathbf{k%
}_{1}^{\prime }}-\omega _{0}+2E_{\mathbf{p}}\right) v_{\mathbf{p}%
}^{4}\left\vert \frac{U\left( \mathbf{p}-\mathbf{k}_{1}^{\prime }\right) }{%
\varepsilon _{\mathbf{k}_{1}^{\prime }}+\zeta _{\mathbf{k}_{1}^{\prime
}}+2E_{\mathbf{p}}+i\delta }\right\vert ^{2}
\end{eqnarray}
where $\zeta_{{\bf k}_{1}^{\prime }}$ is the energy of the core state electron, $\left\langle n_{{\bf k}_{1}^{\prime}}^{f}\right\rangle$ is the occupation of the core electron state, and $U({\bf p}-{\bf k}^\prime_{1})$ is the interaction describing the Auger process. $P_{Aug}^{(2)}$ possesses the same structure as the result shown in Eq.(4) of the main text, albeit with different weighting factors. Note that the calculation of $P_{Aug}^{(2)}$ thus requires knowledge of the detailed momentum and energy structure of the core levels, and that $P_{Aug}^{(2)}$ is only non-zero if there exist occupied core electron states with the same momenta as those of the two photo-electrons probed in the detectors.

\section{Extracting the momentum dependence of the superconducting order parameter in a uniform $d_{x^2-y^2}$-wave superconductor}

\begin{figure}[ht]
\centering
\includegraphics[width=12.5cm]{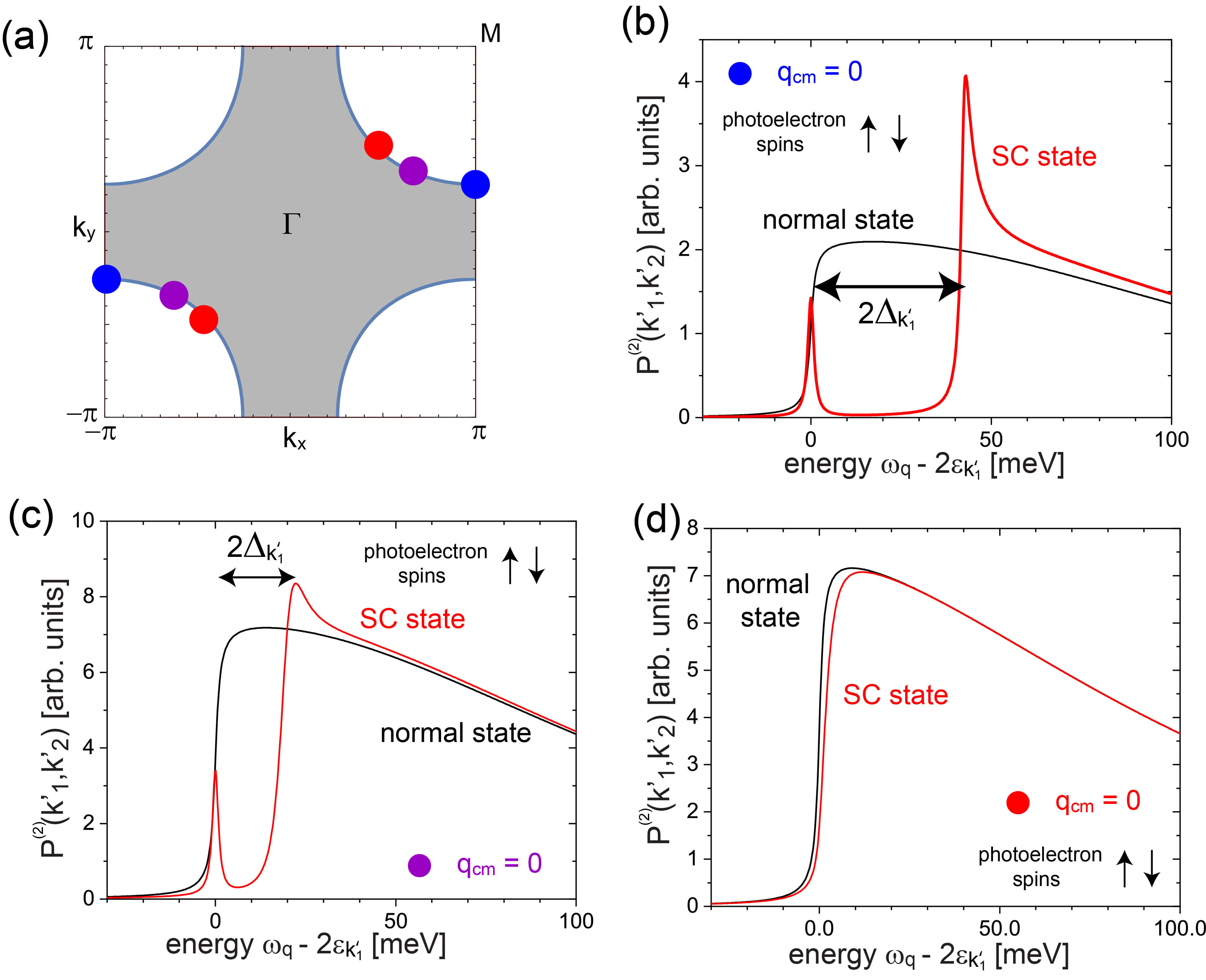}
\caption{(a) Fermi surface of the cuprate superconductors. (b)-(d) $P^{(2)}$ in a uniform $d_{x^2-y^2}$-wave superconductors for two opposite momenta ${\bf k}_1^\prime = -{\bf k}_2^\prime$ along the Fermi surface, as indicated by the sets of blue, purple, and red filled circles.}
\label{fig:SI_Fig1}
\end{figure}
The 2e-ARPES photo-electron counting rate, $P^{(2)}$, can also be employed to map out the momentum dependence of the superconducting gap. To demonstrate this, we consider two photo-electrons with zero center-of-mass momentum, i.e., ${\bf k}_2^\prime = - {\bf k}_1^\prime$, and opposite spin polarization, i.e.,  $\sigma_2^\prime \not =  \sigma_1^\prime$ in a uniform $d_{x^2-y^2}$-wave superconductor.
As shown in Fig.2(b) of the main text for ${\bf k}_{1,2}^\prime$ near the antinodal points, $P^{(2)}$ exhibits a peak at $\Delta \omega = 0$, arising from $P_{SC}^{(2)}$, and a second peak at $\Delta \omega \approx 2 \Delta_{{\bf k}_1^\prime}$ arising from $P_{2cp}^{(2)}$. As ${\bf k}_{1,2}^\prime$ are moved from the antinodal points towards the nodal points (see Fig.~\ref{fig:SI_Fig1}), the second peak remains located at $\Delta \omega \approx 2 \Delta_{{\bf k}_1^\prime}$ and moves down in energy as the superconducting gap decreases. For ${\bf k}_{1,2}^\prime$ at the nodal points, $P^{(2)}$ in the superconducting state is nearly identical to that in the normal state, due to the vanishing superconducting gap. The small differences arise from the fact that the momentum sums in the calculation of $P^{(2)}$ [see Eq.(4) of the main text] probe a small momentum region in the vicinity of the nodal points where the superconducting gap is non-zero, but small.

\section{Effect of a screened Coulomb interaction on the photo-electron counting rate in a uniform $d_{x^2-y^2}$-wave superconductor}
\label{sec:Screening}

\begin{figure}[ht]
\centering
\includegraphics[width=12.5cm]{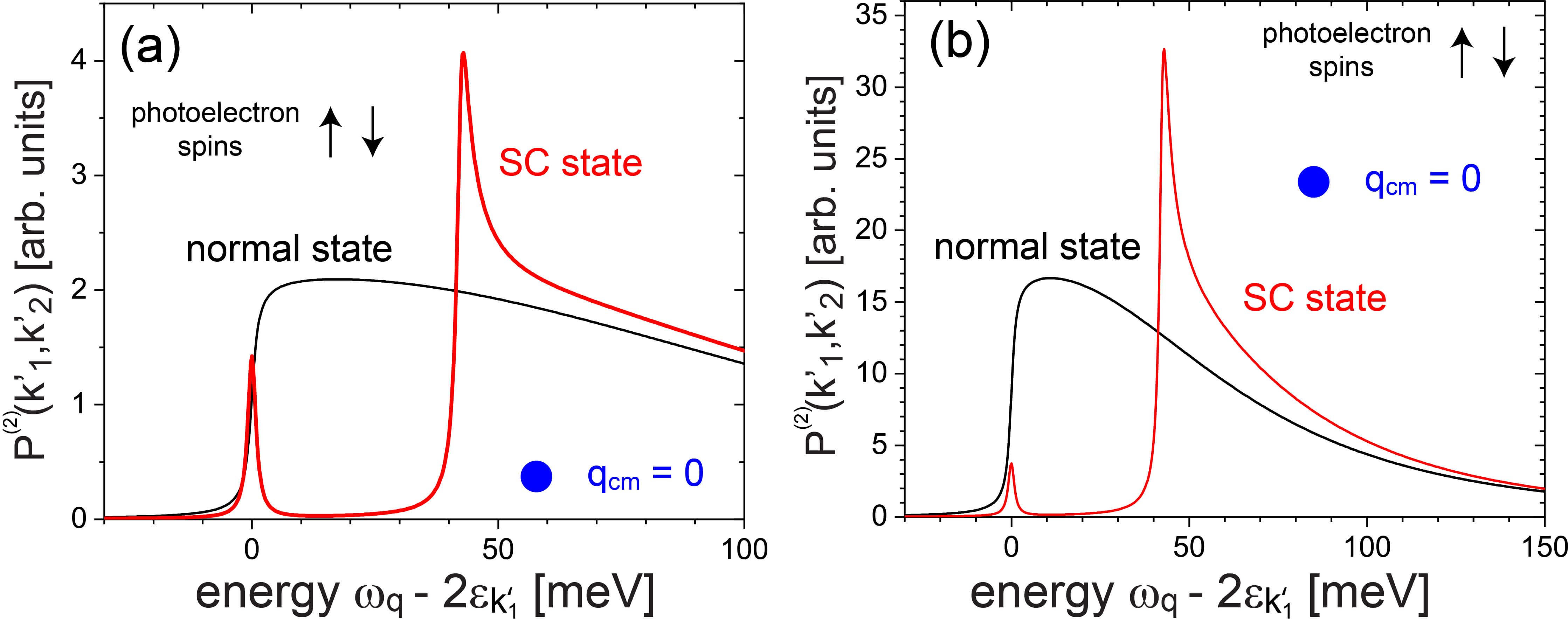}
\caption{$P^{(2)}$ in a uniform $d_{x^2-y^2}$-wave superconductor for two photo-electrons with ${\bf k}_2^\prime = - {\bf k}_1^\prime$, $\sigma_2^\prime \not =  \sigma_1^\prime$, and momenta at the antinodal points [see filled blue circles in Fig.~\ref{fig:SI_Fig1}(a)] and two different screening lengths: (a) $\kappa^{-1} = 10 a_0$, and  (b) $\kappa^{-1} = 20 a_0$.}
\label{fig:SI_Fig2}
\end{figure}

As discussed in the main text, the momentum dependence of the Coulomb interaction is crucial for the observation of a non-zero $P_{SC}^{(2)}$ due to the momentum structure of the superconducting order parameter. One might therefore wonder how the screening of the Coulomb interaction, $V({\bf q}) \sim [{\bf q}^2 + \kappa^{2}]^{-1}$, as reflected in a finite screening length, $\kappa^{-1}$ affects the energy dependence of $P^{(2)}$. To investigate this question, we have computed $P^{(2)}$ for two photo-electrons with zero center-of-mass momentum, i.e., ${\bf k}_2^\prime = - {\bf k}_1^\prime$, and opposite spin polarization, i.e.,  $\sigma_2^\prime \not =  \sigma_1^\prime$, in a uniform $d_{x^2-y^2}$-wave superconductor, for two different screening length, as shown in Fig.~\ref{fig:SI_Fig2}.
While the qualitative structure of $P^{(2)}$ does not change with increasing $\kappa^{-1}$, albeit with an overall intensity increase, we find that the relative height between the peaks arising from $P_{SC}^{(2)}$ and $P_{2cp}^{(2)}$ decreases with increasing $\kappa^{-1}$.